\documentclass[fleqn,usenatbib]{mnras}
 
\usepackage{newtxtext,newtxmath}

\usepackage[T1]{fontenc}

\DeclareRobustCommand{\VAN}[3]{#2}
\let\VANthebibliography\thebibliography
\def\thebibliography{\DeclareRobustCommand{\VAN}[3]{##3}\VANthebibliography}

\usepackage{graphicx}	
\usepackage{amsmath}	
\usepackage[normalem]{ulem}
\usepackage{upgreek}

\newcommand{\simname}[1]{\texttt{#1}}
\newcommand{\bl}[1]{\mbox{\boldmath$ #1 $}}

\newcommand{\Msun}{\,\mathrm{M_{\odot}}}
\newcommand{\Lsun}{\,\mathrm{L_{\odot}}}
\newcommand{\Mdot}{\,\mathrm{M_{\odot}\,yr^{-1}}}
\newcommand{\Mearth}{\,\mathrm{M_{\oplus}}}

\title[Dusty Rings and Outbursts]{Primordial Dusty Rings and Episodic Outbursts in Protoplanetary Discs}

\author[Kadam et al.]{
Kundan Kadam,$^{1}$\thanks{E-mail: kkadam@uwo.ca}
Eduard Vorobyov,$^{2,3}$
and Shantanu Basu$^{1,4}$
\\
$^{1}$Department of Physics and Astronomy, University of Western Ontario, London, Ontario, N6A 3K7, Canada\\
$^{2}$Institute of Astronomy, Russian Academy of Sciences, 48 Pyatnitskaya St., Moscow, 119017, Russia\\
$^{3}$Department of Astrophysics, The University of Vienna, A-1180 Vienna, Austria\\
$^{4}$Institute for Earth \& Space Exploration, University of Western Ontario, London, Ontario, N6A 5B7, Canada
}

\date{Accepted August 24. 2022. Received August 22, 2022; in original form May 6, 2022}

\pubyear{2022}
\graphicspath{{./}{figures/}}

\begin{document}
\label{firstpage}
\pagerange{\pageref{firstpage}--\pageref{lastpage}}
\maketitle

\begin{abstract}
We investigate the formation and evolution of ``primordial'' dusty rings occurring in the inner regions of protoplanetary discs, with the help of long-term, coupled dust-gas, magnetohydrodynamic simulations.
The simulations are global and start from the collapse phase of the parent cloud core, while the dead zone is calculated via an adaptive $\alpha$ formulation by taking into account the local ionization balance.
The evolution of the dusty component includes its growth and back reaction on to the gas.
Previously, using simulations with only a gas component, we showed that dynamical rings form at the inner edge of the dead zone. 
We find that when dust evolution as well as magnetic field evolution in the flux-freezing limit are included, the dusty rings formed are more numerous and span a larger radial extent in the inner disc, while the dead zone is more robust and persists for a much longer time.
We show that these dynamical rings concentrate enough dust mass to become streaming unstable, which should result in rapid planetesimal formation even in the embedded phases of the system.
The episodic outbursts caused by the magnetorotational instability have significant impact on the evolution of the rings. 
The outbursts drain the inner disc of grown dust, however, the period between bursts is sufficiently long for the planetesimal growth via streaming instability.
The dust mass contained within the rings is large enough to ultimately produce planetary systems with the core accretion scenario.
The low mass systems rarely undergo outbursts and thus, the conditions around such stars can be especially conducive for planet formation.
\end{abstract}




\section{Introduction}
\label{sec:intro}

According to the core accretion model of planet formation, the submicron sized dust in a protoplanetary disc (PPD) grows to become pebbles and then to planetesimals that are hundreds of kilometers in size \citep[][]{Pollack+96,Audard+19}. 
The planetesimals are considered to be the building blocks of planets.
Their interactions lead to formation of planetary cores and further processes such as gravitational scattering, accretion of gaseous atmospheres, and planetary migration ultimately result in the large variety of exoplanet systems that are observed \citep{Safronov72, Mizuno80, Lissauer93}. 
The growth of dust from pebble-sized particles to planetesimals faces several theoretical challenges and is not well understood. 
As the dust grains grow to form larger particles, the headwind from the gaseous Keplerian disc causes them to drift towards the maximum in the gas pressure \citep{Whipple72}.
In a simple viscously accreting model of a PPD, the pressure increases monotonically towards the center and the grown dust effectively ends up getting accreted on to the central star \citep{Weidenschilling77b}.
This constitutes the well-known meter-size or drift barrier, which limits the dust size and reduces the dust mass reservoir by draining the disc of grown particles.
The dust growth is also hindered by fragmentation and bouncing barriers; during collisional interactions, the relative velocity between solid particles leads to either fragmentation or bouncing, instead of sticking and growth  \citep{Blum-Wurm2008,Zsom+10}.
On the other hand, several observations suggest that planets should form relatively early during the PPD evolution.
For example, gas giants require the presence of sufficient gas in their surroundings during the runaway accretion phase \citep{BP86,Alibert+05}.
Observations of concentric gaps in young PPDs such as HL Tau can be attributed to the gravitational action of massive planets, indicating early planet formation \citep{Kanagawa+15,Simbulan+17}.
It is also estimated that the class II PPDs do not have enough solid content to form observed populations of exoplanets \citep{Andrews-Williams07,Tychoniec+20}.
These observations imply efficient dust growth during the early stages of the PPD evolution and put severe constraints on the timescale of planet formation.

Substructures in PPDs, such as gaps, rings, spirals, and horseshoe asymmetries, seem to be ubiquitous in the high resolution dust continuum observations by the Atacama Large Millimeter/submillimeter Array \citep[ALMA; e.g.][]{Andrews+09,Andrews+18,Flock+15,Zhang+18,MAPS21}.
The physical origins of the disc substructure can be explained by a variety of mechanisms occurring at a range of distances from the central star and at different stages of PPD evolution. 
In the early stages of disc evolution, spiral structures can form by gravitational instability induced by envelope accretion \citep{VB05, VB06}.
Simulations of magnetohydrodynamic (MHD) disc winds \citep{Suzuki+16,Takahashi-Muto18} and photoevaporative flows \citep{Alexander+14, Ercolano+17} show the formation of rings as well as an inner cavity.
Condensation fronts in the form of snow/ice lines of various volatile species such as water, CO, CO$_2$ and NH$_3$, can lead to enhancements of solids in the form of rings \citep{Stevenson-Lunine88, Cuzzi-Zahnle04, Drazkowska-Alibert17, Molyarova+21}.  
The adsorption of charged species on small dust particles can affect the strength of MRI and produce rings and vortices \citep{Regaly+21}.
Closer to the star, the dust sublimation front at the inner edge of the disc can also lead to dust and gas enhancements \citep{Flock+19}.

{  
Most relevant to our study, observed substructure may also be related to the layered accretion occurring in a PPD.
Canonically, a PPD is thought to evolve viscously and turbulence caused by the magnetorotational instability \citep[MRI;][]{Balbus-Hawley91,Turner+14} is considered to be the primary source of this viscosity.
The MRI acts when the ionized gas in the disk is coupled with its weak magnetic field; the shearing Keplerian motion causes turbulence, resulting in an outward transport of angular momentum.
In the innermost regions of a PPD, thermal ionization of alkali metals makes the disc fully MRI active, while in the outer regions, external sources of ionization such as Galactic cosmic rays penetrate the entire thickness of the disc  \citep{UN81, UN88}.
However, in the intermediate region extending between about 0.5 to 20 au, the typical ionization rates are insufficient to sustain turbulence via MRI. 
This results in accretion occurring only through the surface layers, while a laminar, magnetically dead zone is formed in the midplane \citep{Gammie96}.}
The mismatch of mass transfer rate because of layered accretion in the vicinity of the dead zone can also lead to the formation of rings, horseshoe shapes, and vortices \citep{Dzyurkevich10,RV17}.

The substructures in a PPD essentially form local pressure maxima and can be ideal sites for the accumulation and growth of dust \citep{HB03}. 
When the dust-to-gas ratio is sufficiently high, mechanisms such as streaming instability (SI) can lead to rapid planetesimal formation \citep{YG05,Johansen+07}.
Thus, the aforementioned barriers to dust growth may be overcome at these locations and the disc substructure may be a necessary ingredient in the process of planet formation. 
The forming planets may in turn interact with the gaseous disc and this planet-disc interaction can also shape the disc environment \citep{Kley-Nelson12}.
A gas giant in a PPD opens up a gap during type II migration and the resulting pressure maxima at the gap edge can promote further planetesimal growth \citep{Goldreich-Tremaine80,LP93}.
Innermost regions of such discs, spanning $\lesssim 10$ au, are perhaps most relevant for the formation of Earth-like planets.
However, these regions are difficult to observe directly due to observational constraints on angular size and sensitivity, and are only recently being resolved \citep[e.g.][]{Harter+20,Gravity21}.

During star formation, young stellar objects (YSOs) are observed to undergo sudden and powerful optical outbursts known as FUor-type events \citep{HK96, Audard+14}. 
These outbursts are accretion events that produce luminosities of the order of 100 $\mathrm{L}_\odot$ and last about a hundred years.
A star in thought to undergo an average of about $10$ such eruptions during its formation, accreting nearly one tenth of its final mass through such episodic accretion \citep{Dunham-Vorobyov12}.
Although the phenomenon of episodic accretion appears robust, the origin of these outbursts remains uncertain and there are several candidate mechanisms \citep{Bonnell-Bastien92,Bell-Lin94,Armitage+01,VB05,Kadam+21} 
{  As we shall see later, accumulation of gas in the dead zone region leads to disc instabilities and MRI-outbursts.} 
The outbursts typically occur in the early embedded stages of the disc evolution and thus have the potential to strongly influence the process of planet formation \citep[][]{Abraham+09,Kim+12,Molyarova+18,Vorobyov+20a}.
However, the direct impact of this phenomenon in the context of planetesimal formation is not well understood.

With numerical experiments conducted using hydrodynamics simulations, we showed that during the early stages of PPD evolution, dynamical gaseous rings consistently form in the inner regions at the distance of a few au from the host star \citep{Kadam+19}.
These rings are secularly stable and are caused by viscous torques resulting from spatially nonuniform $\alpha$-parameter near the inner edge of the dead zone.
In this paper, we revisit the idea of ring formation in the inner regions of a PPD, with the inclusion of dynamically coupled dusty component as well as the evolution of magnetic fields in the flux-freezing approximation \citep{VMHD20}.
The treatment of the dead zone is also more complete, with the magnetically active layer calculated using the ionization-recombination balance equation.
These are significant improvements over the earlier simulations that evolved only the gaseous disc with a rather simplistic assumption of constant thickness of the MRI active surface layers \citep{Bae+14}.
With the updated model, we find that ``primordial'' dusty rings indeed form within the innermost few au of the PPD.
These rings are primordial in the sense that they arise in the earliest stages of disc formation due to viscous torques in the magnetically dead zone and should lead to the first generation of planetesimals, especially without the need for a preexisting planet to form pressure traps. 
We update our understanding of these rings by comparative analysis of their structure and evolution with respect to the gas-only simulations. 
We also investigate the prospects of planetesimal formation within the rings via SI and the consequences of episodic accretion.

The structure of the paper is follows. Section \ref{sec:methods} explains our methodology and initial conditions, while the results are presented in Section \ref{sec:results}.
In Section \ref{ssec:structure}, the structure and evolution of the dusty rings is contrasted with the previous model of gas-only disc. 
In Section \ref{ssec:dust}, we focus on the properties of the dusty component in the inner disc, with a focus on planetesimal formation via SI and the effects of episodic outbursts.
In Section \ref{ssec:param}, we present the consequences of varying the cloud core mass and the initial mass-to-flux (mtf) ratio.
We discuss the implications of our findings in Section \ref{sec:discussions} and the conclusions follow in Section \ref{sec:conclusions}.

\section{Methods}
\label{sec:methods}
In this section, we summarize the MHD model that is used for studying the formation and long-term evolution of the PPDs.
The simulations are conducted using the code FEOSAD (Formation and Evolution Of a Star And its circumstellar Disc). 
The latest modifications take into account the coevolution of the coupled dust component as well as magnetic fields in the flux-freezing approximation; the latter consequently leads to a better approximation of the dead zone.
The description in this section belongs primarily to this advanced model, which is described in detail in \cite{VMHD20}.
Note that in Section \ref{sssec:comp}, the inner disc structure is compared with the gas-only simulations, which excluded these two effects \citep{Kadam+19,Kadam+20}. 
The differences between the two models are highlighted where necessary.

The simulations begin with the gravitational collapse phase of a starless cloud core, with the initial profiles of the gas surface density $\Sigma_{\rm g}$ and the angular velocity $\Omega$, consistent with an axisymmetric core collapse
\citep{Basu97}.
The cores with supercritical mtf ratio that can be formed through ambipolar diffusion, have a profile of specific angular momentum versus enclosed mass that remains constant during the collapse.
The evolution continues into the embedded phase of star formation, when the central star is born and is surrounded by a centrifugally supported accretion disc as well as an infalling envelope.
The initial dust-to-gas ratio ($\zeta_{\rm d2g}$) is set to $1\%$, and the dust is in the form of small submicron grains that are fully coupled with the gas. 
The initial core gas temperature is set to $20$ K and a uniform background vertical magnetic field of strength $B_0 = 10^{-5}$ G is assumed.
The spatially uniform mtf ratio, $\lambda \equiv 2\pi \sqrt{G}\Sigma_{\rm g}/B_z$, where $B_z$ is the $z$-component of the magnetic field and $G$ is the gravitational constant, stays constant during the disc evolution in the ideal MHD limit.

The equations of continuity, momentum conservation, and energy transport for the gas component are solved in the thin-disc limit, which can be written as 
\begin{equation}
\label{eq:cont}
\frac{\partial \Sigma_{\rm g}}{\partial t}  + \nabla_p  \cdot \left(\Sigma_{\rm g} \bl{v}_p \right) = 0,  
\end{equation}
\begin{eqnarray}
\label{eq:mom}
\frac{\partial}{\partial t} \left( \Sigma_{\rm g} \bl{v}_p \right) + [\nabla \cdot \left( \Sigma_{\rm g} \bl{v}_p \otimes \bl{v}_p \right)]_p  =   - \nabla_p {\cal P}  + \Sigma_{\rm g} \, \left( \bl{g}_p +\bl{g}_\ast \right) \nonumber \\
+ (\nabla \cdot \mathbf{\Pi})_p  - \Sigma_{\rm d,gr} \bl{f}_p +  \frac{B_z {\bl B}_p^+ }{2 \pi} - H_{\rm g}\, \nabla_p \left(\frac{B_z^2}{4 \pi}\right), ~ ~ ~
\end{eqnarray}
\begin{equation}
\frac{\partial e}{\partial t} +\nabla_p \cdot \left( e \bl{v}_p \right) = -{\cal P}
(\nabla_p \cdot \bl{v}_{p}) -\Lambda +\Gamma + 
\left(\nabla \bl{v}\right)_{pp^\prime}:\Pi_{pp^\prime}, 
\label{eq:energy}
\end{equation}
respectively. Here, the subscripts $p$ and $p^\prime$ refer to the planar components $(r,\phi)$ in polar coordinates, 
$e$ is the internal energy per surface area, $H_{\rm g}$ is the vertical scale height of the gas disc calculated assuming vertical hydrostatic balance, and $\bl{v}_{p}$ is the gas velocity in the disc plane. 
The ideal gas equation of state is used to calculate the vertically integrated gas pressure, ${\cal P}=(\gamma-1) e$ with $\gamma=7/5$.
The term ${\bf g_\ast}$ is the gravitational acceleration by the central protostar, while  
the gravitational acceleration in the disc plane, $\bl{g}_{p}$, takes into account self-gravity of both the gaseous and the dusty components by solving the Poisson equation \citep[see][]{BT08}. 
The term $\bl{f}_p$ is the drag force per unit mass due to the back-reaction of dust on to the gas, and $\Sigma_{\rm d,gr}$ is the surface density of grown dust.
A similar term enters the equation of dust dynamics, as explained later in this section.
The magnetic field inside the disc has only a vertical component $B_z$, while planar components exist at the top and bottom surfaces.
The planar component of the magnetic field at the top surface of the disc is denoted by ${\bl B}_p^+$ and the midplane symmetry is assumed, so that ${\bl B}_p^-=- {\bl B}_p^+$.  
The justification for this approach of representing magnetic fields is discussed in detail in \cite{VB06}, and is based on previously developed evolution equations for a thin magnetized disc \citep{CM93, BM94}. This approximation of the magnetic field has many elements in common with that applied to the calculation of the gravitational field in the thin-disc limit.
The last two terms on the right-hand side of Equation~(\ref{eq:mom}) are the Lorentz force (including the magnetic tension term) and the vertically integrated magnetic pressure gradient.
The vertical component $B_z$ is evolved by solving the ideal MHD form of the magnetic induction equation:
\begin{eqnarray}
\frac{\partial B_z}{\partial t} &=& -\frac{1}{r} \left( \frac{\partial}{\partial r}\left(rv_rB_z\right) + \frac{\partial}{\partial \varphi}\left(v_{\varphi}B_z\right)\right).
\label{eq:Bz}
\end{eqnarray}
The diffusive effects of Ohmic dissipation and ambipolar diffusion are neglected due to high computational cost. 
The planar components of the magnetic field can be written as the gradient of a scalar magnetic potential $\Phi_{\rm M}$ in the external medium. We determine ${\bl B}_p^+$ by solving a Poisson integral with the source term of $(B_z - B_0)/(2 \pi)$, where $B_0$ is the constant background field \citep[see][]{VB06,VMHD20}.

The heating and cooling rates -- $\Gamma$ and $\Lambda$, respectively -- are based on the analytical solution of the radiation transfer equations in the vertical direction \citep{Dong+16}.
The heating function takes into account the irradiation at the disc surface from the stellar as well as background black-body irradiation.
The resulting model has a flared structure, wherein the disc vertical scale height increases with radius.
Both the disc and the envelope receive a fraction of the irradiation energy from the central protostar \citep{VB09}.
The Planck and Rosseland mean opacities are taken from \cite{Semenov+03} and the optical depths in the calculations are proportional to the total dust surface density ($\Sigma_{\rm d,tot}$).
Note that the adopted opacities do not take dust growth into account.

The dust is modelled as consisting of two components; small dust that is coupled with the gas and and grown dust that drifts with respect to the gas and contributes to the back-reaction term \citep{VDust18}. 
More specifically, small dust has a grain size of $a_{\rm min}<a<a_\ast$, and grown dust corresponds to a size of $a_\ast \le a<a_{\rm max}$, where $a_{\rm min}=5\times 10^{-3}$~$\upmu$m, $a_\ast=1.0$~$\upmu$m. Here, $a_{\rm max}$ is a dynamically varying maximum radius of the dust grains, which depends on the efficiency of radial dust drift and the rate of dust growth. 
All dust grains are assumed to have a density of $\rho_{{\rm s}}=3.0\,{\rm g~cm}^{-3}$.
The equations of continuity and momentum conservation for small and grown dust are
\begin{equation}
\label{eq:contDsmall}
\frac{{\partial \Sigma_{\rm d,sm} }}{{\partial t}}  + \nabla_p  \cdot 
\left( \Sigma_{\rm d,sm} \bl{v}_p \right) = - S(a_{\rm max}),  
\end{equation}
\begin{equation}
\label{eq:contDlarge}
\frac{{\partial \Sigma_{\rm d,gr} }}{{\partial t}}  + \nabla_p  \cdot 
\left( \Sigma_{\rm d,gr} \bl{u}_p \right) = S(a_{\rm max}),  
\end{equation}
\begin{eqnarray}
\label{eq:momDlarge}
\frac{\partial}{\partial t} \left( \Sigma_{\rm d,gr} \bl{u}_p \right) +  [\nabla \cdot \left( \Sigma_{\rm
d,gr} \bl{u}_p \otimes \bl{u}_p \right)]_p  &=&   \Sigma_{\rm d,gr} \, \left( \bl{g}_p + \bl{g}_\ast \right) + \nonumber \\
 + \Sigma_{\rm d,gr} \bl{f}_p + S(a_{\rm max}) \bl{v}_p,
\end{eqnarray}
where $\Sigma_{\rm d,sm}$ and $\Sigma_{\rm d,gr}$ are the surface densities of small and grown dust, respectively. 
The term $\bl{u}_p$ describes the planar components of the grown dust velocity, and $S(a_{\rm max})$ is the rate of conversion from small to grown dust per unit surface area, which is a function of the maximum size of the dust. 
The dust is assumed to mix vertically with the gas, which is a reasonable approximation for a young disc evolving under gravitational and viscous torques.
The rate of small to grown dust conversion $S(a_{\rm max})$ is derived from the assumption that the size distributions of both the dust populations can be described by a power law with a fixed exponent of $-3.5$.  
The discontinuity in the dust size distribution at $a_*$ is assumed to get smoothed out due to the $S$ term, which implies the dominant role of dust growth rather than dust flow \citep[see for details][]{Molyarova+21}.
The evolution of the maximum size that the grown dust can achieve is expressed as a continuity equation of the form
\begin{equation}
\frac{\partial a_{\rm max}} {\partial t} + ({\bl u}_{\rm p} \cdot \nabla_p ) a_{\rm max} = \cal{D},
\label{eq:dustA}
\end{equation}
where
\begin{equation}
\cal{D}=\frac{\rho_{\rm d} {\it v}_{\rm rel}}{\rho_{\rm s}}
\label{eq:GrowthRateD}
\end{equation}
is the growth rate that accounts for coagulation, in which $\rho_{\rm d}$ is the total dust volume density.
{  The relative dust-to-dust velocity, $v_{\rm rel}$, is calculated by considering the main sources of relative velocities -- the Brownian motion and turbulence \citep[for details, see][]{VDust18}.}
The maximum size that the dust can achieve is limited by the fragmentation barrier
\begin{equation}
 a_{\rm frag}=\frac{2\Sigma_{\rm g}v_{\rm frag}^2}{3\pi\rho_{\rm s}\alpha c_{\rm s}^2},
 \label{afrag}
\end{equation}
where $v_{\rm frag}$ is the fragmentation velocity and $c_{\rm s}$ is the local sound speed \citep{Birnstiel+12}. 
{  
Laboratory experiments as well as collision simulations of aggregates give a value between a fraction of ${\rm m\,s^{-1}}$ to a few tens of ${\rm m\,s^{-1}}$ for fragmentation velocities \citep{Wada+13,Guttler+10,Kelling+14}.
This mainly depends on the grain composition, e.g. ice grains may be able to grow at significantly higher velocities.
In our simulations $v_{\rm frag}$ is set to a conservative value of 3 ${\rm m\,s^{-1}}$ \citep{VMHD20}.}
The growth rate $\cal{D}$ is set to zero when $a_{\rm max}$ exceeds $a_{\rm frag}$.
Note that in our approach, the dust growth is limited to keep the size of dust particles within the Epstein regime. 
Hence the back-reaction term is expressed as
\begin{equation}
     \bl{f}_p = \frac{\bl{v}_p-\bl{u}_p }{t_{\rm stop}}, 
      \label{eq:f_p}
\end{equation}
where the stopping time is $t_{\rm stop} = a_{\rm max} \rho_{\rm s}/ c_s \rho_{\rm g}$, and is calculated using an asymptotic approximation \citep{Stoyanovskaya+18,Stoyanovskaya+20}.

The viscosity required for the mass and angular momentum transport in a PPD is considered primarily due to the turbulence generated by MRI \citep{Balbus-Hawley91,Turner+14}. Note that the gravitational torques are another major mass and angular momentum transport mechanism in young disks, which are considered self-consistently in FEOSAD \citep{VB09}.
MRI acts in the presence of magnetic field and causes turbulence in the weakly ionized gas in the shearing Keplerian disc.
The primary sources of ionization, such as the galactic cosmic rays, are external to the disc and penetrate the disc from both the surfaces.
{ The gas surface density in the inner regions of the disc is typically large, and outside of the innermost 0.5 au the midplane temperature is not high enough for thermal ionization of alkali metals.
Thus, this region is insufficiently ionized and the disc accretes through a magnetically layered structure,} wherein most accretion occurs via the MRI-active surface layers and a magnetically dead zone is formed at the midplane \citep{Gammie96}.
In FEOSAD, the viscosity is taken into account via the viscous stress tensor $\Pi$ in Eq. (\ref{eq:mom}). 
The kinematic viscosity is parameterized using the \cite{SS73} $\alpha$ prescription. 
In order to mimic the accretion through a layered disc, we consider an effective and adaptive parameter $\alpha_{\rm eff}$ as a weighted average \citep{Bae+14,Kadam+19}: 
\begin{equation}
\label{eq:alphaeff}
    \alpha_{\rm eff} = \frac{\Sigma_{{\rm MRI}} \, \alpha_{{\rm MRI}} + \Sigma_{{\rm dz}} \, \alpha_{{\rm dz}}}{\Sigma_{{\rm MRI}} + \Sigma_{{\rm dz}}},
\end{equation}
where $\Sigma_{{\rm MRI}}$ is the gas column density of the MRI-active layer and $\Sigma_{{\rm dz}}$ is that of the magnetically dead layer at a given position, so that $\Sigma_{\rm g} = \Sigma_{{\rm MRI}} +\Sigma_{{\rm dz}}$. Here, $\alpha_{\rm MRI}$ and $\alpha_{{\rm dz}}$ correspond to the strength of the turbulence in the MRI-active layer and the dead zone, respectively. 
In the simulations, $\alpha_{\rm MRI}$ is set to the canonical value of $0.01$.
{  However, according to recent 3D MHD simulations of MRI bursts in PPDs, the effective $\alpha$ values in the inner disc region directly involved in the outbursts can be much larger \citep[e.g.][]{Zhu+20}.
The value of $\alpha_{\rm MRI}$ is thus set to 0.1 within the inner 5 au of the disc.
This larger value of the $\alpha$ parameter gives a better fit to the duration of FU Orionis-type outbursts inferred from observations \citep{VMHD20}.} 
The 5 au boundary for this transition is adopted because the MRI outbursts are typically contained within the innermost 2-3 au.
The abrupt drop in $\alpha_{\rm MRI}$ should not influence the disk evolution, since the dead zone in this region remains robust and $\alpha_{\rm eff}$ remains near its minimum value at all times (see sections \ref{ssec:structure} and \ref{sssec:MRIburst}).
Due to the nonzero residual viscosity arising from hydrodynamic turbulence driven by the Maxwell stress in the active layer, a small but nonzero value of $10^{-5}$ is considered for $\alpha_{{\rm dz}}$ \citep{OH11}.

Canonically, the thickness of active layer is considered to be a constant, at about $100\, {\rm g\,cm^{-2}}$, which is the average penetration depth of cosmic rays \citep{UN81}. 
FEOSAD is upgraded to calculate $\Sigma_{\rm MRI}$ from the ionization fraction ($x$), which in turn is determined from the ionization balance equation
\begin{equation}
(1-x)\xi = \alpha_{\rm{r}} x^2n_{\rm{n}} + \alpha_{\rm{d}} xn_{\rm{n}},
\label{eq:ion}
\end{equation}
where $\xi$ is the ionization rate, $\alpha_{\rm{r}}$ is radiative recombination rate, $n_{\rm n}$ is the number density of neutrals, and $\alpha_{\rm{d}}$ is the total rate of recombination on to the dust grains \citep{DS87,DK14}.
While calculating ionization rate, the ionization by cosmic rays and radionuclides is considered.
{  The ionization due to the stellar far-ultraviolet and X-ray radiation is neglected because at median T Tauri luminosities, their penetration depth is about $10\, {\rm g\,cm^{-2}}$, significantly smaller than that of the cosmic rays \citep{Bergin07, Cleeves13, IG99}.}
The total recombination rate for each dust population (small and grown) is calculated as
\begin{equation}
\langle \alpha_{{\rm d}} \rangle = \langle X_{{\rm d}} \cdot \sigma_{{\rm d}} \cdot v_{{\rm i}} \rangle,
\label{eq:recomb}
\end{equation}
where $X_{{\rm d}}$ is the dust-to-gas volume number density ratio in the disc midplane, $\sigma_{{\rm d}} = \pi a^2$ is the grain cross-section.
The quantity $v_{{\rm i}}$ is the approximate thermal speed of ions with mass $30\,m_{{\rm H}}$, representing the dominant ionic species, e.g. HCO$^{+}$, N$_2$H$^+$. 
The degree of thermal ionization is obtained from the ionization of potassium as a metal with the lowest ionization potential and is added to the ionization rate obtained from Eq. (\ref{eq:ion}).
The cosmic abundance of potassium is set to $10^{-7}$ for these calculations.

Although the coevolution of the magnetic field is implemented in the flux-freezing limit, a hybrid approach is considered for calculating the thickness of the MRI active surface layer associated with the dead zone.
The wavelength of the most unstable MRI mode is $\lambda_{{\rm crit}}= 2\pi \eta /v_{{\rm a}}$, where $\eta$ is the total magnetic diffusivity and $v_{{\rm a}}$ is the Alfv{\' e}n speed.
The column density of the active layer $\Sigma_{{\rm crit}}$ is obtained by equating $\lambda_{{\rm crit}}$ to the gas scale height of the disc, yielding
\begin{equation}
    \Sigma_{{\rm crit}} = \left[\left(\frac{\pi}{2}\right)^{1/4}\frac{c^2m_{\rm e}\langle\sigma v\rangle_{\rm en}}{e^2}\right]^{-2}B_z^2 H_{\rm g}^3 x^2 \, ,
    \label{eq:DZ}
\end{equation}
where $e$ is the charge of an electron, $m_{\rm e}$ is the mass of an electron and $\langle\sigma v\rangle_{\rm{en}}= 10^{-7}\,\mbox{cm}^3\,\mbox{s}^{-1}$ is the slowing-down coefficient \citep{Nakano84}.
Note that in this approach, the effects of Ohmic diffusivity, which dominate the inner disc region, are considered, and other nonideal MHD effects are neglected \citep{Balbus-Terquem01,Kunz-Balbus04}.
When the dead zone is present, $\Sigma_{\rm g} > 2 \times \Sigma_{\rm crit}$ and $\Sigma_{\rm MRI}$ is set to $\Sigma_{\rm crit}$.
The factor of two comes from the two surfaces of the disc.
If $\Sigma_{\rm g} \leq 2 \times \Sigma_{\rm crit}$, then the disc is fully MRI active ($\Sigma_{\rm MRI} = \Sigma_{\rm g}$) and there is no dead zone.

\begin{table*}
\caption{List of Simulations}
\label{table:sims}
\begin{tabular}{|l|}
\hline
\begin{tabular}{p{2cm}p{1.8cm}p{1.1cm}p{1.2cm}p{1cm}p{1.2cm}p{1.2cm}p{2.5cm}}
 Model Name &  $ M_{ \rm core} (\Msun) $  &  $\beta$  &   $ \Omega_0 $ &  $r_{\rm in}$&  $r_{\rm out}$ &  $\lambda$ &   Salient differences \\
\end{tabular}\\ \hline
\begin{tabular}{l}
$\kern-\nulldelimiterspace\left.
\begin{tabular}{p{2cm}p{1.5cm}p{1.2cm}p{1.2cm}p{1cm}p{1.2cm}p{1.2cm}p{2.5cm}}
\simname{model-1}   &  1.45  & 0.23 &  3.0 &  0.52 & 0.035 & 2.0 \\
\simname{model-2}   &  0.83  & 0.23 &  5.2 &  0.52 & 0.02 & 2.0 \\
\simname{model-2L}   &  0.83  & 0.23 &  5.2 &  0.52 & 0.02 & 10.0 \\
\simname{model-3}   &  0.21  & 0.23 &  20.8 &  0.52 & 0.005 & 2.0 \\
\end{tabular}\right\}$ Gas and dust dynamics, MHD$^*$
\\ 
\end{tabular}  
\\ \hline
\begin{tabular}{l}
$\kern-\nulldelimiterspace\left.
\begin{tabular}{p{2cm}p{1.5cm}p{1.2cm}p{1.2cm}p{1cm}p{1.2cm}p{1.2cm}p{2.5cm}}
\simname{model-2G}   &  0.83  & 0.23 &  5.2 &  0.52 & 0.02 & 2.0  \\ 
  \end{tabular}\right\}$ Gas dynamics
\end{tabular}  
\\
\hline
\end{tabular}\\
\raggedright $^*$ MHD in flux-freezing approximation with updated dead zone calculations. 
\end{table*}

The inner boundary of the computational domain needs special attention, as the processes taking place in its vicinity may be important to the dynamical evolution of the entire disc \citep{Vorobyov+19}. 
The inner boundary cannot be placed near the stellar surface where magnetospheric accretion takes place, while it cannot be placed much farther, as important dynamical processes such as the MRI outbursts may be entirely missed.
In our simulations, the inner boundary is placed at 0.52 au from the disc center; this relaxes the Courant condition and also captures the episodic accretion during the PPD evolution \citep{CFL28}.
Another complication arises from the type of boundary condition used. 
The gas flow near the boundary may be wave-like with spiral features, and a typical outflow-only condition results in an artificial drop in the gas surface density.
To overcome this issue, we use a special inflow?outflow boundary condition at the inner edge of the disc, wherein the matter is allowed to flow freely from the disc to the sink cell and vice versa \citep{VDust18,Kadam+19}.
The mass exiting the inner boundary is divided between the sink cell and star, with the ratio set to 95:5. 
The material in the sink cell may enter back into the immediate active cell in the disc, based on its radial velocity and surface density gradient.
The flow of matter also carries magnetic flux, hence, the inner boundary condition also modifies $B_z$ based on the amount of magnetic flux transported.
The inner boundary maintains the initial spatially constant mtf ratio ($\lambda$) across time, which serves as a test of disc evolution in the ideal MHD limit.
The inner boundary conditions also conserve mass and magnetic flux budget of the system.
Thus, the behaviour of gas and dust in the vicinity of the inner boundary can be trusted to be free from any numerical artefacts.
The outer boundary is standard free outflow, where the material is only allowed to leave the computational domain.

In this study, we present the results from five simulations as listed in Table \ref{table:sims}.
The first four simulations are conducted with the inclusion of dust dynamics as well as MHD effects in the flux-freezing approximation as described in this section. This also includes updated calculations of the dead zone. For this dust+MHD approach, \simname{model-2} is designated as the fiducial and representative simulation.
For comparison with earlier results, a corresponding simulation was conducted neglecting these two effects.
This is the fiducial gas-only simulation--\simname{model-2G}.
More specifically, Eqs. \ref{eq:Bz}-\ref{eq:DZ} were not solved, with the exception of Eq. \ref{eq:alphaeff}, where the $\alpha_{\rm eff}$ was based on a constant $\Sigma_{\rm MRI} =100\, {\rm g~cm^{-2}}$.
The thermal ionization threshold due to alkali metals was set at a constant temperature of 1300 K, over which the disc was considered fully MRI-active \citep{Umebayashi83,Gammie96}. 
The two fiducial models are otherwise identical with an initial gas mass of $0.83 \Msun$ for their cloud core.
The remaining models are similar to those presented in \cite{VMHD20} and the naming convention reflects this fact. 
The effects of varying the mass is explored with \simname{model-1} and \simname{model-3}, which have an initial cloud core gas mass of 1.45 and $0.21 \Msun$, respectively.
A greater value of mtf ratio, $\lambda=10$, is considered in \simname{model-2L}.
For consistency, all models have an identical ratio of initial rotational to gravitational energy $\beta$. 
The grid used is cylindrical in the plane of the disc, with with logarithmic spacing in the radial direction and uniform spacing in the azimuthal direction.
The numerical resolution of all simulations is $256\times256$, so that the highest grid resolution in the radial direction of 0.018 au is in the innermost cell located at 0.52 au.

\section{Results}
\label{sec:results}

\subsection{Structure and Evolution of the Rings}
\label{ssec:structure}
In this section, we describe the structure and evolution of the rings formed in the inner regions of a PPD.
The disc structure of the fiducial dust+MHD simulation (\simname{model-2}) is compared with that of the fiducial gas-only simulation (\simname{model-2G}). 
Table \ref{table:sims} shows that these two models are evolved from identical initial conditions, with the cloud core gas mass of 0.83 ${\rm M}_\odot$.
As described in Section \ref{sec:methods}, the major difference between the two models is the coevolution of the coupled dust component and the magnetic field in the ideal MHD regime.
The comparison is done for the first 0.7 Myr and the system at the end of this time can be considered as a T Tauri object (see Section \ref{sssec:MRIburst} for class definitions).
In \cite{Kadam+19}, we showed that concentric gaseous rings form in the inner regions of the disc at a few au from the central star due to viscous torques in gas-only simulations. 
The motivation behind this comparison between the two fiducial simulations is to update our understanding of PPDs after incorporating more advanced physics in the disc evolution model.

\begin{figure*}
\centering
  \includegraphics[width=18cm]{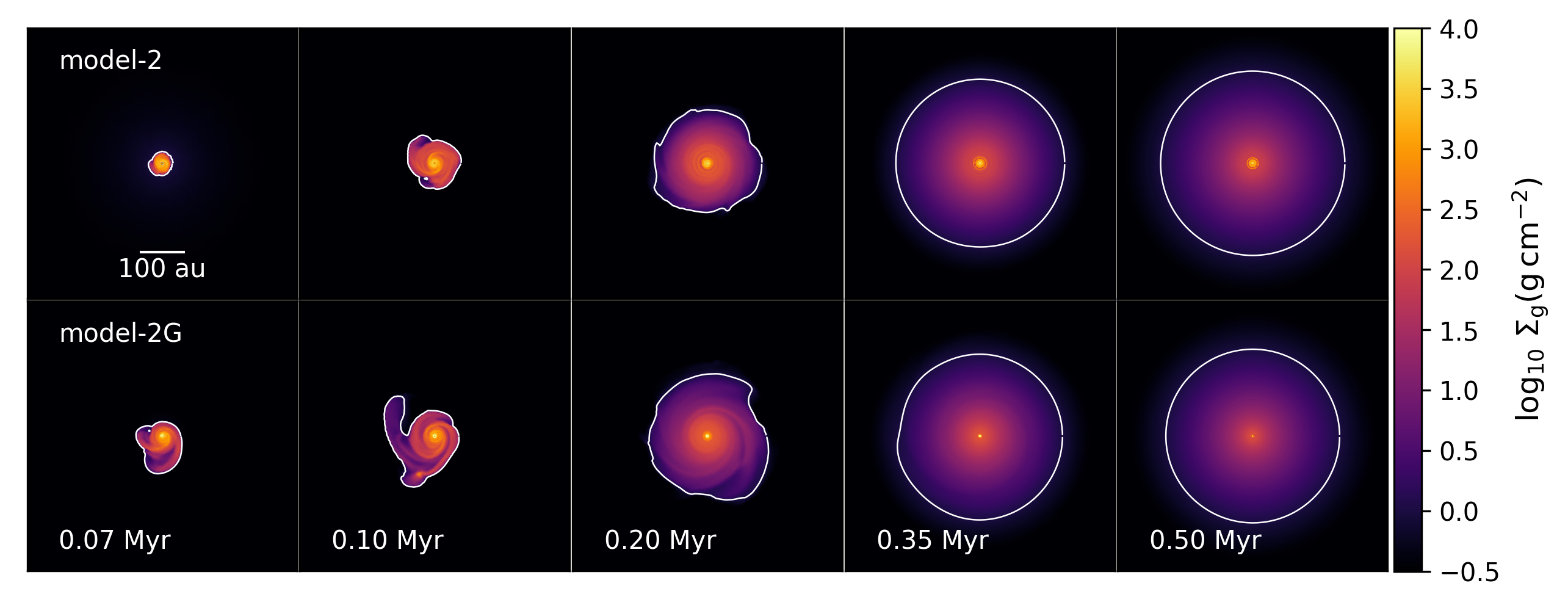}
\caption{Evolution of the disc gas surface density distribution for the fiducial dust+MHD model-- \simname{model-2} (top row)--and corresponding layered disc model with gas evolution only--\simname{model-2G} (bottom row)--over a region of 500 x 500 au. The white contour shows the approximate extent of the disc with $\Sigma_{\rm g} = 1 {\rm \, g\, cm^{-2}}$. } 
\label{fig:global}
\end{figure*}

Figure \ref{fig:global} shows the global picture, where the evolution of the gas surface density of the two fiducial models is depicted over the inner $500\times 500$ au region.
The time is measured from the beginning of the simulation, when the cloud core starts its collapse.
The overall evolution of the two discs looks similar at this scale. 
The early phases are dominated by gravitational instability (GI) when the disc is relatively massive as compared to the central protostar.
The irregular spiral structures and formation of clumps are observed in both the simulations.
As each disc evolves, the viscous spread of the circumstellar disc is observed, and they also become azimuthally uniform and smooth.
The disc formation in the case of \simname{model-2} is delayed by a short time, about 20 kyr, due to the additional support provided by the magnetic field.
Although \simname{model-2G} shows an initially larger disc at a given time, the difference becomes insignificant after 0.35 Myr.
The major difference between the two models at this scale is that \simname{model-2} shows a consistently greater gas surface density in the innermost regions as compared to \simname{model-2G}.

\begin{figure*}
\hspace{1cm}  Azimuthal profiles    \hspace{3.8cm} \simname{model-2G}  \hspace{2.8cm}  \simname{model-2}\\
\begin{tabular}{c}
\vspace{-0.19cm}\hspace{0.30cm}\includegraphics[width=0.977\textwidth]{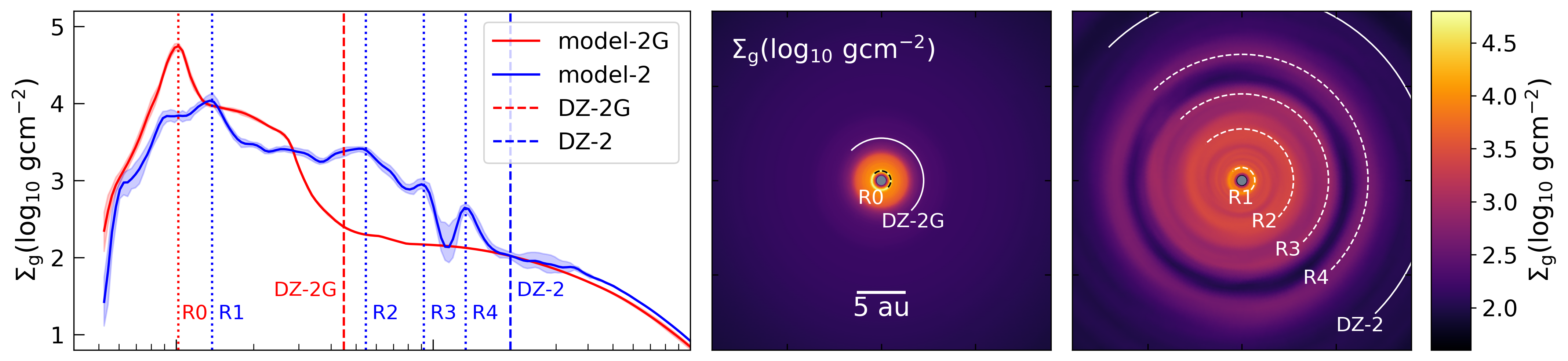} \\ 
\vspace{-0.19cm}\hspace{0.135cm}\includegraphics[width=0.985\textwidth]{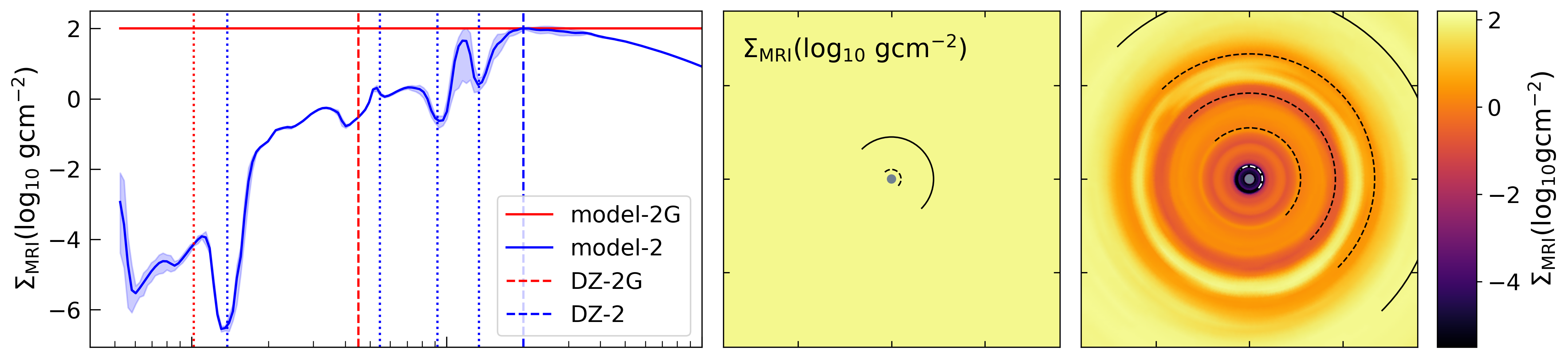} \\
\vspace{-0.19cm}\hspace{0.12cm}\includegraphics[width=\textwidth]{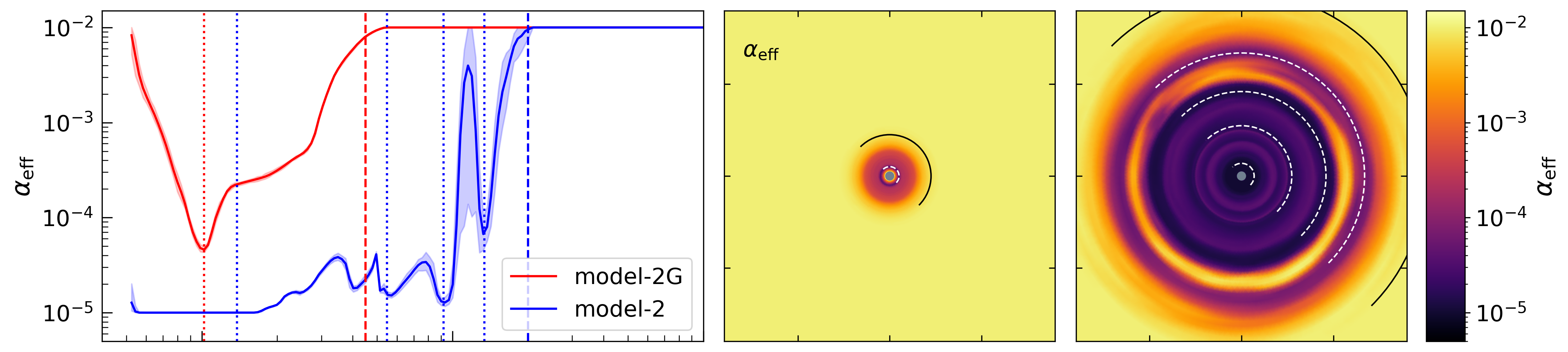} \\
\vspace{-0.25cm}\hspace{-0.02cm}\includegraphics[width=\textwidth]{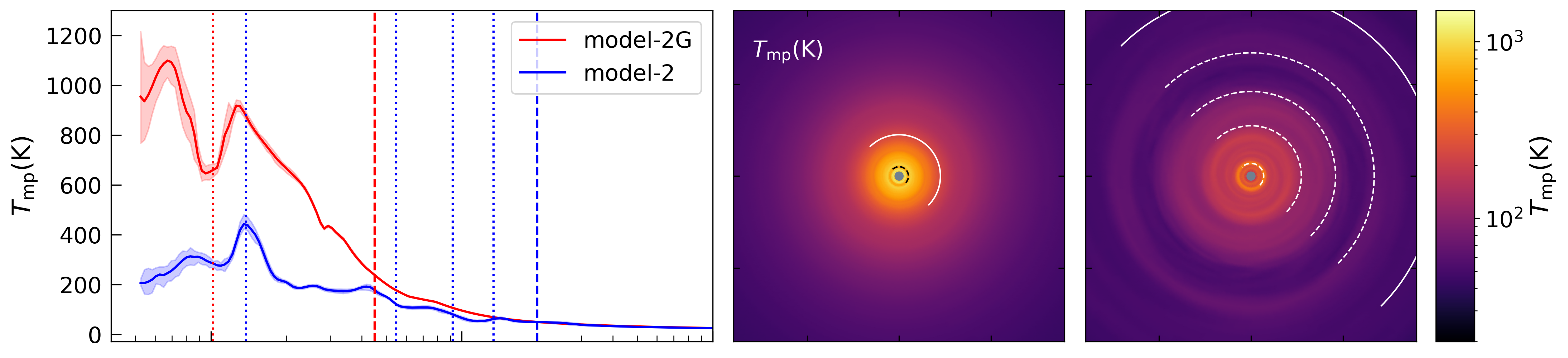} \\
\hspace{0.2cm}\includegraphics[width=0.985\textwidth]{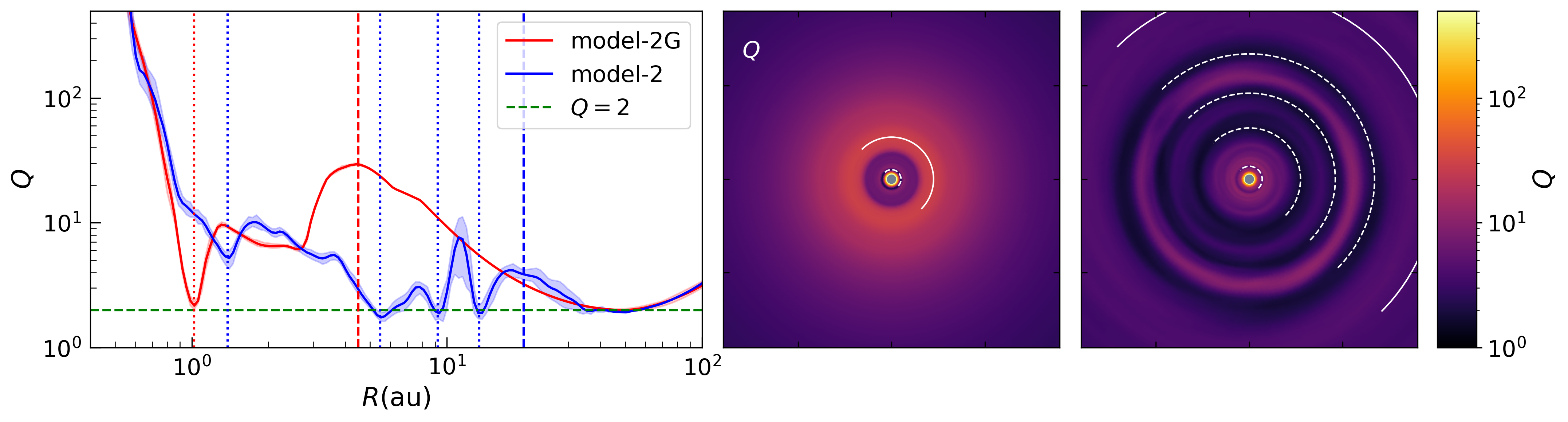} \\
\end{tabular}
\caption{Typical inner disc structure of the fiducial gas-only disc (\simname{model-2G}) is compared against that of the dust+MHD disc (\simname{model-2}) at 0.35 Myr, with respect to quantities--gas surface density, $\Sigma_{\rm MRI}$, $\alpha_{\rm eff}$, midplane temperature, and Toomre's $Q$-parameter. 
Left: Profiles of the azimuthally averaged parameters, with the shaded area showing the extent between the maximum and the minimum value at the given radius in the inner 100 au region. The vertical dashed lines mark the outer extent of the dead zones for the two models (DZ-2G and DZ-2), while the dotted lines marks the location of the rings (R0-R4). 
The green horizontal dashed line marks the $Q=2$ level.
Right: The distribution of the same parameters is plotted in the inner $36 \times 36$ au box, where the dead zones and rings are shown in solid and dashed lines, respectively.}
\label{fig:inner1}
\end{figure*}

In Figure \ref{fig:inner1}, we focus on the innermost regions of the two fiducial disc simulations (\simname{model-2} and \simname{model-2G}).
These stationary snapshots correspond to 0.35 Myr and capture typical disc structure.
The global disc structure at this time can be seen in the fourth column of Figure \ref{fig:global} for both the discs. 
At this time, most of the GI induced spirals have died down and the discs have become increasingly axisymmetric.
Figure \ref{fig:inner1} compares the two models with respect to five quantities -- the gas surface density, the $\alpha_{\rm eff}$ as given by Eq. (\ref{eq:alphaeff}), the thickness of the MRI active layer, the midplane temperature $T_{\rm mp}$, and Toomre's $Q$-parameter, respectively. 
Here we calculate its value using a modified definition
\begin{equation}
Q=\frac{c_{\rm s}\Omega}{\pi G(\Sigma_{\rm g}+\Sigma_{\rm d,sm}+\Sigma_{\rm d,gr})},    
\end{equation}
which includes the contribution by the dust component \citep{VDust18}. 
Note that in the ideal MHD limit, there is an effective value $Q_{\rm eff} = Q(1+2\lambda^{-2})^{1/2}/(1-\lambda^{-2})$ for a thin sheet and instability occurs when $Q_{\rm eff} \lesssim 1$ \citep[][]{Das2021}. 
Given the uncertainties in the exact instability threshold value when dealing with averaged quantities, we do not take into account the order unity correction and use the instability criterion of $Q \lesssim 1$.
In each row of Figure \ref{fig:inner1}, the plot on the left hand side compares the azimuthally-averaged quantities in the inner 100 au.
The shaded region shows the extent between the maximum and minimum quantity at the given radius.
The boxes on the right compare the 2D distributions of the corresponding variables in the inner $36 \times 36$ au box.
The extent of the dead zone for both the simulations is defined as the region wherein $\Sigma_{\rm g} \geq 2\times \Sigma_{\rm MRI}$.
The dead zone extends much farther in \simname{model-2} ($\approx$ 11 au), as compared to \simname{model-2G} ($\approx$ 4 au).
The azimuthal profiles of the quantities from the two simulations tend to converge outside of the dead zone for \simname{model-2}.
The regions identified as rings in terms of gas surface density are marked in Figure \ref{fig:inner1} by vertical dotted lines in the azimuthally averaged plots and by dashed lines in the 2D distributions.
The gaseous rings do not always coincide with the dusty rings and we elaborate on this later in this section.

The first row of Figure \ref{fig:inner1} depicts the gas surface density for the two fiducial models.
At this time, \simname{model-2G} shows formation of axisymmetric ring, labelled R0 near the inner edge of the dead zone at 1 au.
Note that \simname{model-2G} has a smaller initial gas mass of the cloud core ($0.83 \Msun$) as compared to the models presented in \cite{Kadam+19} ($1.13 \Msun$). 
Thus, due to the limited mass reservoir, only a single ring is formed in this case of \simname{model-2G}.
In contrast, in \simname{model-2} multiple rings are formed throughout the extent of the dead zone, including near its outer edge.
These rings are labeled R1-R4 and are located at 1.1, 5.4, 9.5, and 11 au, respectively.
Note that the exact number and locations of these rings are subject to change as the disc evolves in time, however, the general differences between the inner disc structure of the two simulations remain consistent. 

The second row of Figure \ref{fig:inner1} shows the thickness of the MRI active layer for the two models.
For \simname{model-2G}, the value of $\Sigma_{\rm MRI}$ is set to a constant value of $100\, {\rm \,g\, cm^{-2}}$.
For the dust+MRI \simname{model-2}, $\Sigma_{\rm MRI}$ is calculated by solving ionization-recombination balance equation and thus, it represents a region characterized by small ionization fraction and effective diffusion of the magnetic field (see Eq. \ref{eq:DZ}).
The $\Sigma_{\rm MRI}$ calculated this way is several orders of magnitude smaller than the canonical value. 
The variable $\Sigma_{\rm MRI}$ in \simname{model-2} is one of the key differences in the two simulations.

The third row of Figure \ref{fig:inner1} shows the depth of the dead zones in terms of $\alpha_{\rm eff}$ (see Eq. \ref{eq:alphaeff}).
The differences in the azimuthal profiles of $\alpha_{\rm eff}$ for the two simulations are significant and central in understanding the characteristic structure and evolution of the discs.
For \simname{model-2G}, $\alpha_{\rm eff}$ reaches a minimum value of $\approx 10^{-4}$ for a narrow annulus and remains above this value within most of the extent of the dead zone, while it also shows gradual increase at the outer edge of the dead zone.
In contrast, $\alpha_{\rm eff}$ for \simname{model-2} remains at $\approx 10^{-5}$ throughout the dead zone and it increases relatively sharply at its outer boundary.
As the $\alpha_{\rm eff}$ consistently remains at the minimum possible value, the dead zone can be considered as much more robust and less prone to change due to variations in the disc conditions.
The width of the outer transitional region in \simname{model-2} is about 5 au, which is consistent with that expected from detailed calculations including microphysics \citep{Dzyurkevich13}.
These differences in the structure of the dead zone for the two simulations have significant consequences for the inner disc and essentially explain the formation of the multiple rings in the dust+MHD models.
The spatially non-uniform $\alpha$-parameter results in a variable mass accretion rate in the disc and its low value implies a severely narrow bottleneck for angular momentum transport.
This leads to a global accumulation and enhancement of gas as it moves inward from the fully turbulent outer regions of the disc. 
Additionally, the expression for viscous torque has a component directly proportional to the negative of the gradient of kinematic viscosity.
As a consequence, the material is accelerated outward at the inner boundary and it is accelerated inwards at the outer boundary of the dead zone \citep{Kadam+19}.
For \simname{model-2}, the abrupt drop in $\alpha_{\rm eff}$ would cause a significant inward torque at the outer boundary of the dead zone.
The dust in the disc is typically attracted towards the pressure maxima, however, a pressure bump can also induce over-density of particles due to a ``traffic jam'' caused by variations of logarithmic pressure gradient \citep{Carrera+21}.
This can lead to a mismatch between gas and dust accumulation and explain the rings R2 and R3, which are not associated with robust pressure maxima.
Thus, the simulations show that the dead zone is not a uniform structure with enhanced gas surface density, but it has complex substructure in the form of a multitude of dynamically evolving rings.

The fourth row in Figure \ref{fig:inner1} shows the temperature structure of the discs. 
The midplane temperature of a canonical $\alpha$-disc decreases monotonically, since both viscous heating ($Q_{\rm vis} \propto \Sigma_{\rm g} \nu \Omega^2$) and stellar irradiation diminish with increasing radial distance.
The low value of $\alpha_{\rm eff}$ in the dead zone results in low viscosity and the decreases viscous heating in this region, which in turn lowers the midplane temperature.
Thus, \simname{model-2} shows a consistently lower midplane temperature than \simname{model-2G} because of the lower viscosity associated with its dead zone. 
Consider the temperature near the location of the rings.
The gaseous ring R0 in \simname{model-2G} shows a lower temperature in its vicinity because of the associated lesser viscous heating.
However, the dust accumulated in the rings of \simname{model-2} causes inefficient cooling due to the increased opacity.
Thus, the temperature in the vicinity of the rings in \simname{model-2} can be larger, e.g. ring R1.
The aspect ratio of the disc is given by $H_{\rm g}/R=c_{\rm s}/v_{\rm k}$, where $H_{\rm g}$ is the gas scale height, $c_{\rm s}$ is the local sound speed, and $v_{\rm k}$ is the Keplerian velocity.
Since the $T_{\rm mp} \propto c_{\rm s}^2$, with the inclusion of dust physics in \simname{model-2}, the rings show larger scale heights than their surroundings.
However, this increase in $T_{\rm mp}$ as well as $H_{\rm g}$ is limited to the innermost rings, where the disc heating is viscously dominated.
In the regions beyond a few au, the rings show lower $T_{\rm mp}$ and $H_{\rm g}$, similar to the gas-only \simname{model-2}.
Note that the changes in the scale height are marginal and do not cause significant shadowing effects.

The last row of Figure \ref{fig:inner1} shows the Toomre's $Q$-parameter for the two fiducial models.
The $Q$-parameter quantifies the importance of the self-gravity, as spiral patterns and clumps start appearing in the disc as its value approaches unity. 
For \simname{model-2G}, the $Q$-parameter approaches a low value in the vicinity of the gas ring due to its high surface density.
The \simname{model-2} shows similar structure, however, only the rings near the outer dead zone edge approach the $Q=2$ level. 
In conclusion, the robustness and extent of the dead zone increases in our more advanced model, which also forms more numerous rings.

\begin{figure*}
\hspace{-2cm} \simname{model-2G} \hspace{7cm} \simname{model-2}  \\
\begin{tabular}{l}
\hspace{-0.56cm}  \vspace{-0.2cm}\includegraphics[width=\textwidth]{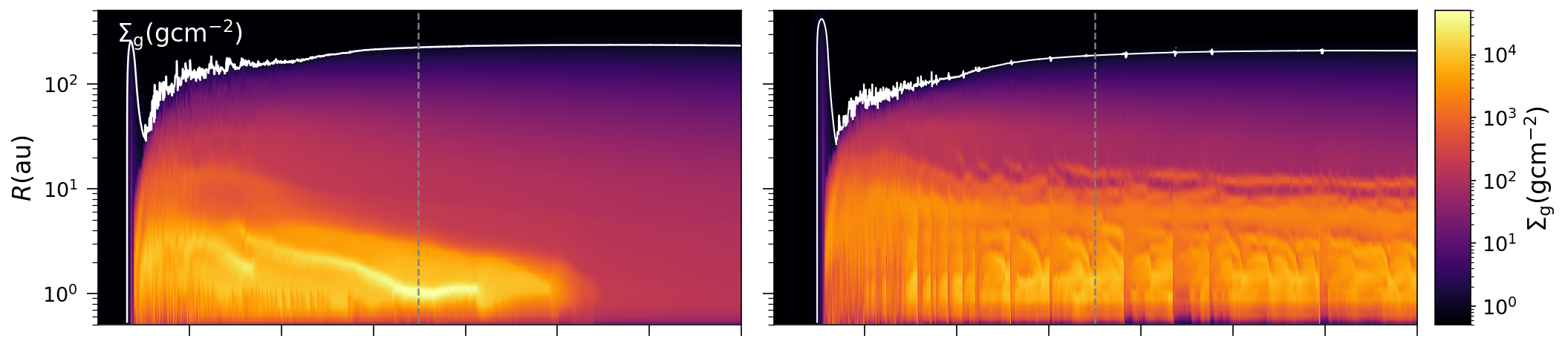} \\
\hspace{-0.56cm}  \vspace{-0.2cm}\includegraphics[width=1.005\textwidth]{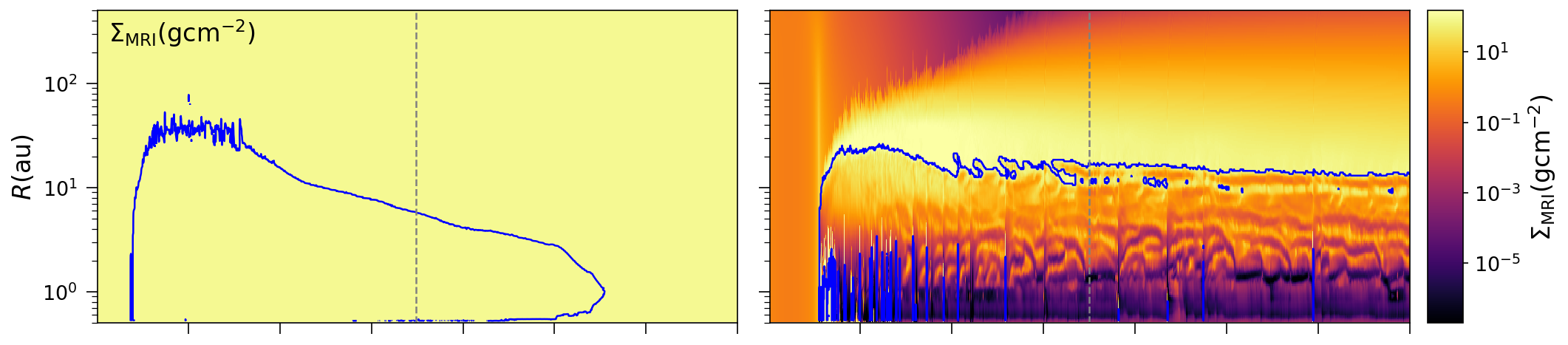} \\ 
\hspace{-0.55cm} \vspace{-0.2cm}\includegraphics[width=\textwidth]{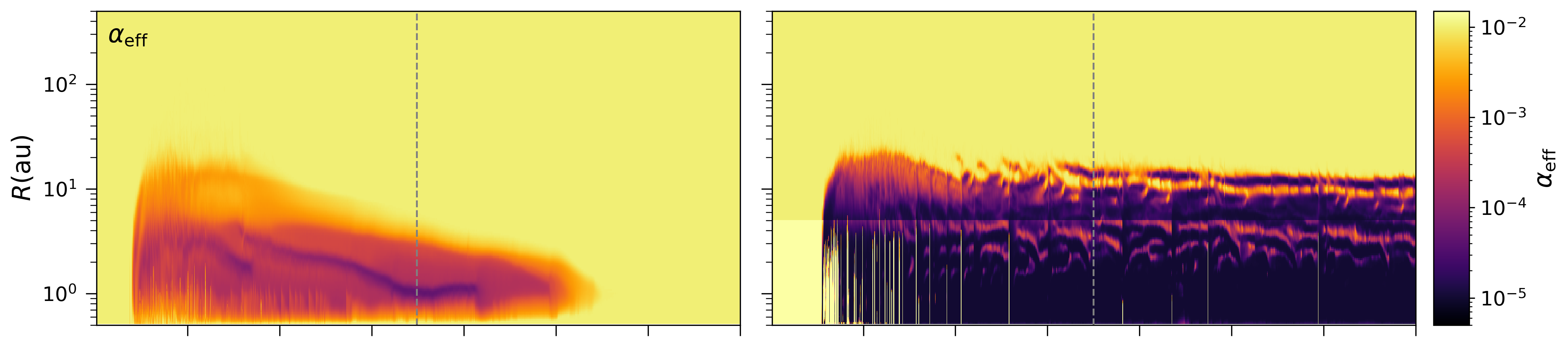} \\
 \hspace{-0.55cm} \vspace{-0.2cm} \includegraphics[width=\textwidth]{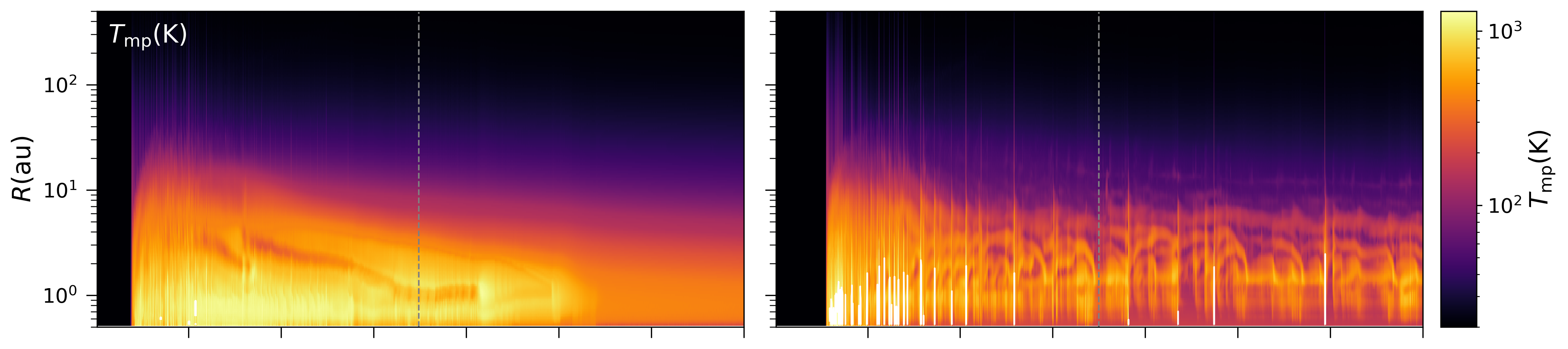} \\  
\hspace{-0.55cm} \vspace{-0.2cm}\includegraphics[width=\textwidth]{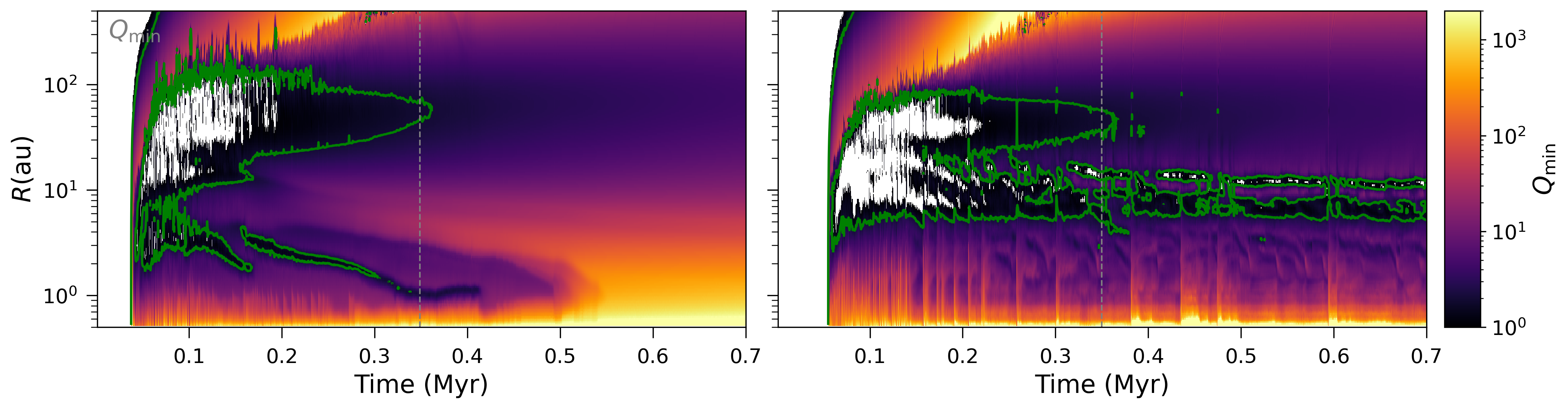} \\
\end{tabular}
\caption{Spacetime plots for the two fiducial models--\simname{model-2G} and \simname{model-2}--with the rows depicting the evolution of azimuthally averaged quantities--$\Sigma_{\rm g}$, $\Sigma_{\rm MRI}$, $\alpha_{\rm eff}$, $T_{\rm mp}$, and the minimum in $Q$-parameter. 
The white curves in $\Sigma_{\rm g}$ shows 1 ${\rm g\, cm^{-2}}$ contours, while the blue contours in $\Sigma_{\rm MRI}$ shows the extent of the dead zone.
The white lines in the midplane temperature plots show $T_{\rm crit}=1300$ K contours. 
The white shaded area in $Q$-parameter plot shows gravitationally unstable regions where $Q\leq1$, while the green contours show the $Q=2$ level.
The vertical dashed lines show the time corresponding to the static images in Figure \ref{fig:inner1}.
The dust+MHD \simname{model-2} shows qualitatively different behaviour as compared to the gas-only \simname{model-2G}. 
}
\label{fig:spacetime1}
\end{figure*}

So far, we described typical ring structures occurring inside the dead zone.
However, they correspond to a static snapshot of the inner disc that is highly dynamical.
In order to compare the disc behaviour across time, we plot the spacetime diagrams of some of the characteristic variables in Figure \ref{fig:spacetime1}.
The figure compares the azimuthally averaged quantities -- $\Sigma_{\rm g}$, $\Sigma_{\rm MRI}$, $\alpha_{\rm eff}$, $T_{\rm mp}$, and the azimuthal minimum in the $Q$-parameter.
Each row shows the evolution of the variables for the gas-only simulation \simname{model-2G} on the left and dust+MHD \simname{model-2} on the right.
Note that the ordinates in these plots are in logarithmic units. 
The temporal resolution of the azimuthally-averaged profiles is set by the output frequency of the data, which is 1000 yr in our numerical code. 
The resolution of the $Q$-parameter is 2000 yr, since the minimum value is calculated using 2D output files.
The snapshots in Figure \ref{fig:inner1} (and later Figure \ref{fig:inner2}) correspond to 0.35 Myr, which is marked by the vertical dashed line at 0.35 Myr.
The general disc structure outside of about 20 au, which is approximately the extent of the dead zone in \simname{model-2}, looks similar for both the simulations throughout the 0.7 Myr.

Consider the first row in Figure \ref{fig:spacetime1}, which depicts the evolution of gas surface density. 
The white contour plotted at $\Sigma_{\rm g} = 1 {\rm \,g\, cm^{-2}}$ marks the approximate outer extent of the PPD.
The blue contour in the next row marks the boundary of the dead zone, i.e. $\Sigma_{\rm g} = 2 \times \Sigma_{\rm MRI}$.
As expected, \simname{model-2G} forms long-lived, gaseous rings in the inner disc region. 
The density enhancements in the inner disc begin soon after the disc formation, while axisymmetric structures resembling rings start appearing after about 0.15 Myr.
The prominent ring diminishes around 0.42 Myr, while the outer extent of the dead zone slowly decreases as the disc evolves and it completely disappears at 0.55 Myr.
Note that the total initial gas mass of \simname{model-2G} is much less than in the fiducial simulation presented in \cite{Kadam+19}.
This essentially results in less complicated ring structure, with only one ring appearing at a time, while the dead zone also lasts for a relatively short duration.
On the other hand, the dead zone is about 20 au wide in \simname{model-2} and its extent only slowly decreases with time.
The structures within are complicated and we will see later that they are more clearly identified as axisymmetric rings in the dust plots.

The second row of Figure \ref{fig:spacetime1} compares the thickness of the MRI-active layer of the two models. 
For \simname{model-2}, even with all the sources of ionization considered, the resultant ionization fraction is of the order of $10^{-14}$ in the dead zone and reaches $10^{-16}$ at the location of the rings. These values are insufficient to sustain a significant thickness of the MRI active layer (see Equation \ref{eq:DZ}).
The $\Sigma_{\rm MRI}$ values are thus several orders of magnitude less than the canonically assumed value of 100 ${\rm g\, cm^{-2}}$ for \simname{model-2G}.
The third row shows $\alpha_{\rm eff}$ and hence reflects the viscous structure of the disc.
The region outside of the dead zone is fully MRI active.
The depth of the dead zone in terms of $\alpha_{\rm eff}$ is much deeper in the case of \simname{model-2}, with values hovering near the minimum of $10^{-5}$.
These differences can be attributed to the aforementioned treatment of the calculation of $\Sigma_{\rm MRI}$. 
As a result of the low viscosity in the dead zone, the rings formed are stable for a correspondingly longer viscous timescale.

The fourth row in Figure \ref{fig:spacetime1} shows the temperature evolution of the disc.
As seen earlier with the static frames, the disc in \simname{model-2} is much cooler as compared to \simname{model-2G} due to the decrease in viscous heating.
The white contours in this plot mark the critical temperature of 1300 K, above which the gas in \simname{model-2G} is considered fully MRI active due to ionization of the alkali metals.
The sharp discontinuities or spikes in \simname{model-2} near its inner boundary result from from MRI bursts that punctuate and violently disturb the inner disc structure.
Each burst can be clearly observed in temperature as well as in all of the spacetime plots of $\Sigma_{\rm g}$ and $\alpha_{\rm eff}$.
{  These MRI bursts occur throughout the evolution of \simname{model-2} and their origin as well as implications for the disc evolution are discussed in detail in Sections \ref{sssec:plt} and \ref{sssec:MRIburst}.}
However, such bursts are diminished in magnitude and are limited to only the earliest times for \simname{model-2G}.
Note that for this gas-only model, the occurrence of bursts depends on the adopted value of parameter $T_{\rm crit}$ \citep{Bae+13,Kadam+20},
and \simname{model-2} is free from this uncertainty.
Additionally, the phenomenon of MRI bursts has a lower mass limit in terms of the mass of the parent cloud core and the disc around a star, because of the lower available mass reservoir.
In \cite{Kadam+19}, simulations with total initial gas mass of 1.152$\Msun$ underwent bursts and those with $0.346 \Msun$ did not show outbursts.
The exact mass threshold of this transition in episodic accretion is not well known and it may be gradual and dependent on details of the modeled physics.
We think that the newer dust+MHD simulations are more reliable and should better model the PPD evolution.

The last row of Figure \ref{fig:spacetime1} shows the evolution of the azimuthal minimum in the Toomre's $Q$-parameter. A minimum is chosen over an azimuthal average because the $Q$-parameter quantifies clumping due to the disc self-gravity, which is a local phenomenon.
The initial prestellar collapse phase can be seen before the formation of the disc in each case, where the cloud is gravitationally unstable at all radii.
As expected, discs in both the models are gravitationally unstable at early times and at large distances.
{  In addition to viscous torques, GI plays a role in transporting the material from the outer disc towards the dead zone \citep{VB09}.}
However, this initial phase of gravitational instability extends noticeably inward and lasts for a longer duration for \simname{model-2}.
Small fragments of white can be seen in the vicinity of the outer edge of the dead zone in \simname{model-2} throughout the evolution of the disc. This suggests that the rings in these regions show brief instances of GI fragmentation. 
Such GI fragmentation does not occur near the rings in \simname{model-2G}.
The GI fragments in the inner disc can lead to generation of large scale spirals, which can efficiently carry angular momentum through a large region of the disc for a brief period of time \citep{Kadam+19}.
{ This may constitute a major source of angular momentum transport in the dead zone of \simname{model-2}, which has significantly reduced viscosity.}

\subsection{Dust properties}
\label{ssec:dust}
\subsubsection{Comparison with gas-only model}
\label{sssec:comp}

\begin{figure*}
\hspace{3.5cm}  Azimuthal profiles    \hspace{3.2cm} \simname{model-2G}  \hspace{2.8cm}  \simname{model2}\\
\begin{tabular}{l}
\vspace{-0.22cm}\includegraphics[width=\textwidth]{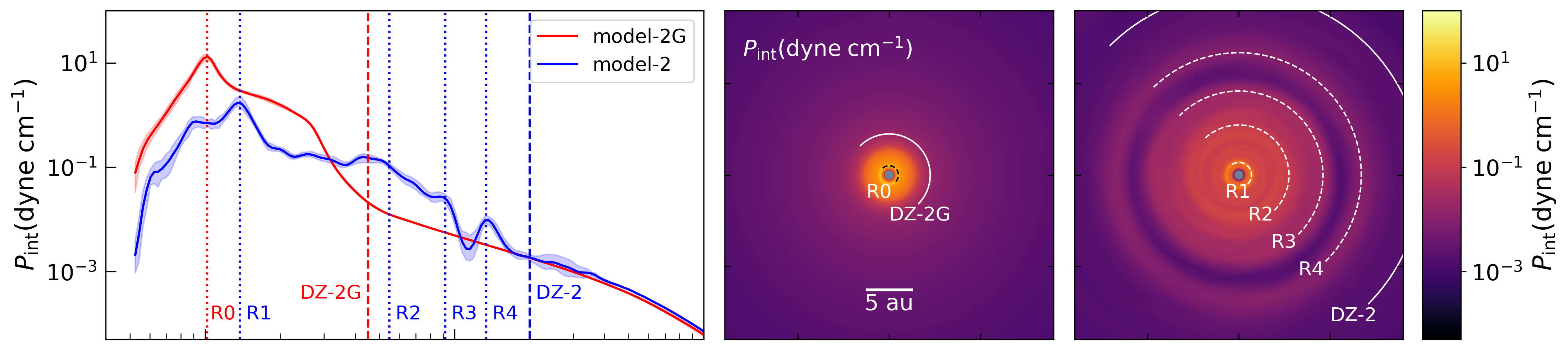} \\ 
\vspace{-0.22cm}\hspace{0.2cm}\includegraphics[width=0.98\textwidth]{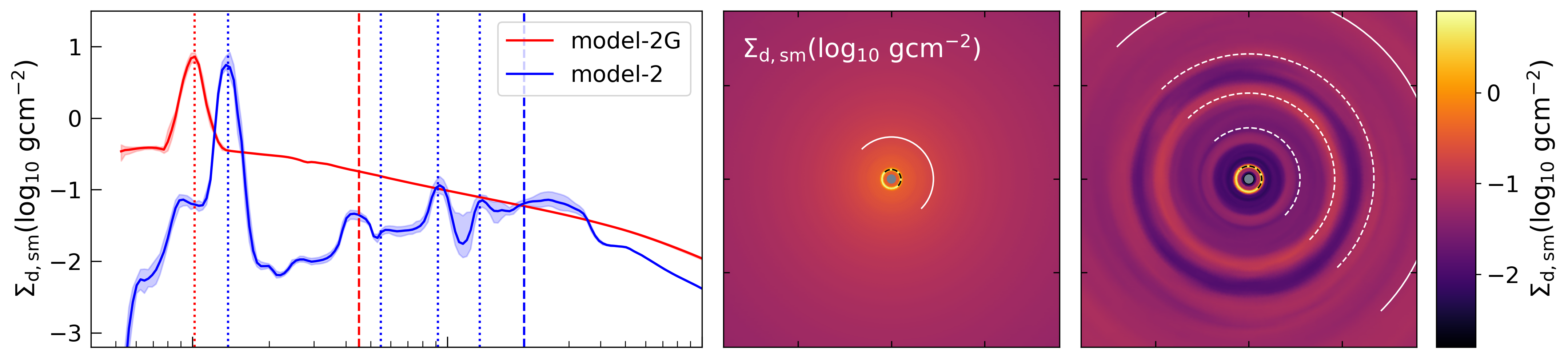} \\ 
\vspace{-0.22cm}\hspace{0.2cm}\includegraphics[width=0.97\textwidth]{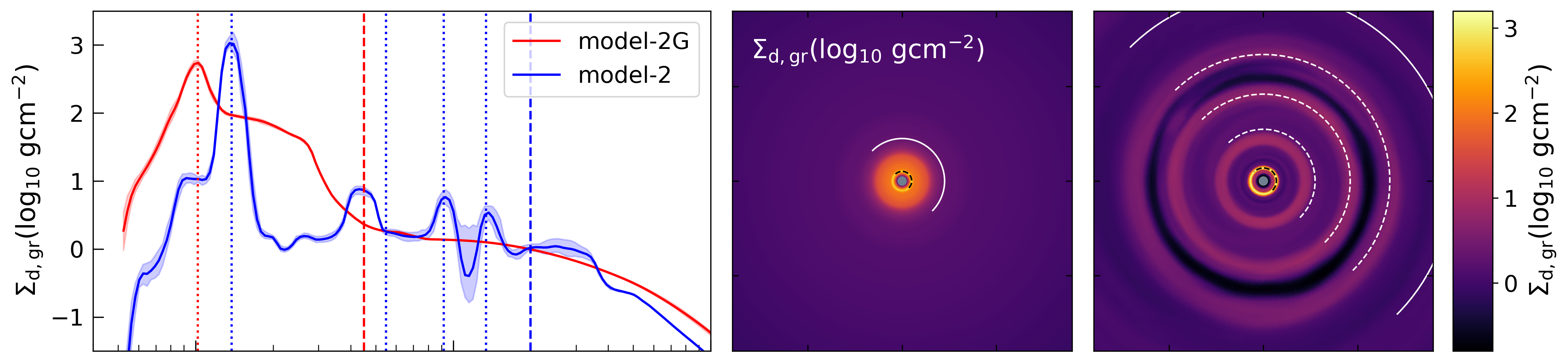} \\ 
\vspace{-0.22cm}\hspace{0.05cm}\includegraphics[width=\textwidth]{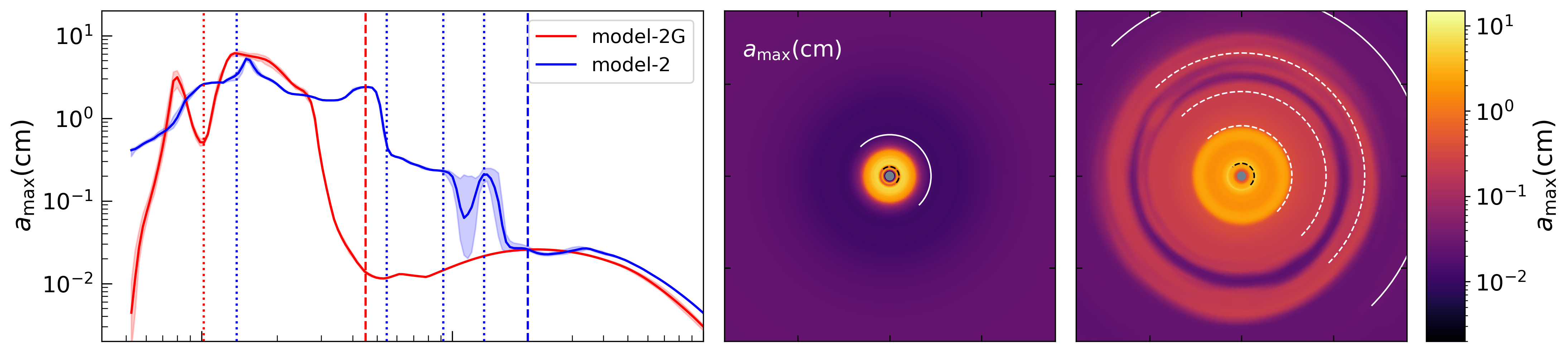} \\ 
\hspace{0.05cm}\vspace{-0.22cm}\includegraphics[width=\textwidth]{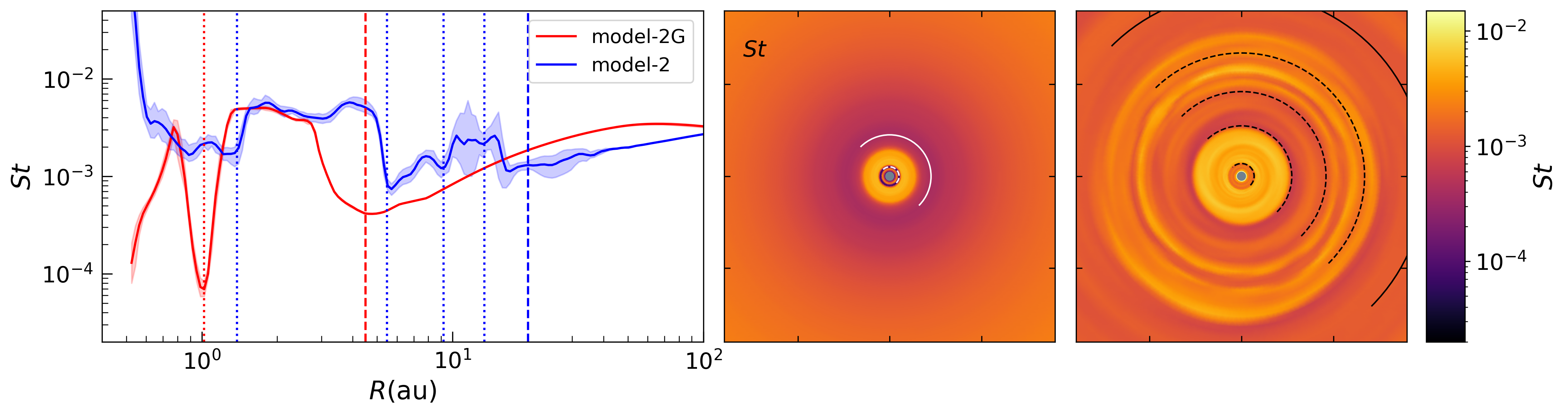}
\end{tabular}
\caption{Typical inner disc structure of the fiducial gas-only disc (\simname{model-2G}) is compared against that of the dust+MHD disc (\simname{model-2}) at 0.35 Myr, with respect to quantities related to dust properties--integrated pressure, small and grown dust surface densities, fragmentation size of the dust grains, and Stokes number.
The legends are similar to Figure \ref{fig:inner1}, with the lines marking the gaseous rings (R0-R4) and the outer extents of the dead zone (DZ-1, DZ-2).  
}
\label{fig:inner2}
\end{figure*}

In this section, we investigate the inner disc structure with respect to the accumulation of dust and its growth in the two fiducial models. 
Since \simname{model-2G} did not have any dust physics, some simplistic assumptions were made for the purpose of comparison between the two fiducial models.
Consider Figure \ref{fig:inner2}, which compares typical inner disc structure of the fiducial gas-only disc (\simname{model-2G}) against that of the dust+MHD disc (\simname{model-2}).
Consider the vertically integrated pressure in the first row of Figure \ref{fig:inner2}.
The pressure maxima within PPDs are especially important for the process of planet formation.
As the dust particles within a Keplerian disc are attracted towards pressure maxima, these locations effectively function as dust traps and potential sites for planetesimal formation.
For \simname{model-2G}, there is only one pressure maximum throughout the disc at the location of the ring R0. 
On the other hand, three out of four rings in \simname{model-2} show associated pressure maxima.
The disc as a whole is dynamically evolving and ring R2 seems to be caused by a traffic jam mechanism and hence shows a non-typical structure.
The new dust+MHD simulations suggest that there are several dust traps available for dust growth throughout the dead zone that may aid planet formation over a large range of radii.

The second and third rows of Figure \ref{fig:inner2} show the surface density of small and grown dust respectively. For \simname{model-2}, the small dust is submicron sized initial dust reservoir and is gradually converted in grown dust as the disc forms and evolves.
The continuity equation is solved for the small dust that is coupled with the gas, while the dynamics of grown dust are controlled by friction with the gas component and by gravitational interactions with the total system (see Section \ref{sec:methods}).
The \simname{model-2G} does not include any dust physics and its dust content is estimated as follows.
The total dust surface density is assumed to be 1\% of the gas surface density.
The partition between small and grown dust is calculated in a similar way to \simname{model-2}, by assuming the boundary between the two components at 1 $\upmu$m and a power-law distribution with an exponent of $-3.5$.   
This gives us an order of magnitude estimate of the the conditions in the gas-only simulation and puts the widely used assumption of a constant dust-to-gas ratio into perspective.
Consider the azimuthal profiles of the dusty components in second and third row of Figure \ref{fig:inner1}.
For both the models, most of the dust mass is in the grown component, as it is one to two orders of magnitude larger than the small dust.
For \simname{model-2}, the small dust is progressively converted into grown dust and hence it is depleted throughout the disc.
This depletion can be noticed in the outermost parts of the disk, where the the two models show similar profiles, while they differ significantly within the extent of the dead zone.
For \simname{model-2}, both small and grown dust is concentrated much more in the inner ring, R1, as compared to the outer rings.
The ring R2 observed in gas surface density does not coincide with any of the dusty rings.
However, both small and grown dust is concentrated at a short distance inside R2, confirming the hypothesis that it is a structure undergoing dynamical changes due to dust-gas interactions described earlier.
The dust in \simname{model-2} shows much more intricate and complex structures, as compared to relatively smooth profiles for \simname{model-2G}.
Note that the observations of resolved discs similarly show that the dusty rings do not necessarily coincide with analogous structures in gas tracers \citep[e.g.][]{Isella+16,Yen+16}.

Dust depleted rings and cavities are often observed in the high angular resolution images of PPDs obtained in dust 
continuum emission at submillimeter wavelength \citep{Long+18,Nienke+19}, while similar substructures are also seen in gas with the CO isotopologues and other tracers \citep{Huang+18,SP+20}. 
The depletion of dust in between the rings is significant in \simname{model-2}.
We calculate the dust depletion factor in between the rings as 
$\delta_{\rm d} = \Sigma_{\rm d,min}/\Sigma_{\rm d,avg}$, where $\Sigma_{\rm d,min}$ is the minimum value of the total dust surface density in between two rings. Here, $\Sigma_{\rm d,avg}$ is a proxy for the background dust surface density, calculated as half of the average value of the two maxima that encompass this minimum. 
The dust depletion factors between the rings depicted in Figure \ref{fig:inner2} are $\delta_{\rm d, R1-R2}= 0.016$, $\delta_{\rm d, R2-R3}= 0.27$, and $\delta_{\rm d, R3-R4}= 0.73$.
With an analogous definition of gas depletion factor between the rings, we find that $\delta_{\rm g, R1-R2}= 0.5$, $\delta_{\rm g, R2-R3}= 0.88$, and $\delta_{\rm g, R3-R4}= 0.38$ at the same instance of time depicted in Figure \ref{fig:inner1}.
Thus, the depletion in the dust is much stronger than that in the gas.
With the increase in radius, the depletion factor in dust rings with respect to their surroundings typically becomes less pronounced and approaches unity.
This trend is not strictly observed in gas, which is expected as it shows significant variations near the sharp outer boundary of the dead zone.

The fourth row of Figure \ref{fig:inner2} shows the dynamically varying maximum radius of dust grains, $a_{\rm max}$, for \simname{model-2}.
Fragmentation is a process wherein the dust particles break into smaller fragments due to collisions, which tends to put an upper limit to the dust size, especially in the inner disc \citep{Birnstiel+12,VDust18}.
However, in our hydrodynamic model, the term for the drag force between dust and gas components ($\bl{f}_p$ in Eq. \ref{eq:momDlarge}) is valid only in the Epstein regime.
In order to maintain the validity of our model, we limit the dust growth to a size $a_{\rm r}= 9/4 \ell$, where $\ell= m_{\rm H_2} \sqrt{2 \pi}  H_{\rm g} / (7 \times 10^{-16} \Sigma_{\rm g})$ 
is the mean free path of the hydrogen molecule. Here, $m_{\rm H_2}$ is mass of the hydrogen molecule and $H_{\rm g}$ is gas scale height.
Similar to the dust+MHD models, the $a_{\rm max}$ in \simname{model-2G} is calculated as the minimum between $a_{\rm frag}$ and $a_{\rm r}$, although note that the dust physics as well as the drift barrier are understandably neglected.
As seen in Figure \ref{fig:inner2}, the dust aggregates achieved a maximum size of a few centimeters in the majority of the dead zone for both the models, while the sizes were very similar in the outer disc.
The larger dust grain size was extended over a much larger region in the case of \simname{model-2}.
The maximum in $a_{\rm max}$ is usually coincident with the ring locations. 
Although these maxima may deviate from the rings, as $a_{\rm max}$ is related to $\Sigma_{\rm g}$, $\alpha_{\rm eff}$ and the midplane temperature through $c_{\rm s}$, and these quantities evolve in a rather complicated way.

The last row of Figure \ref{fig:inner2} compares the Stokes numbers for the two fiducial models calculated for the corresponding $a_{\rm max}$.
The Stokes number quantifies the aerodynamic properties of the dust and its coupling with the surrounding gas.
The Stokes number for a particle of size $a$, under typical assumptions, is 
\begin{equation}
    {\rm St} = \frac{\Omega_{\rm k} \rho_{\rm s} a}{\rho_{\rm g} c_{\rm s}},
\end{equation}
where $\Omega_{\rm k}$ is the Keplerian angular velocity, $\rho_{\rm g}$ is the midplane volume density of the gas, and $a$ is the particle size.
For both the models, the Stokes number stays well below unity throughout the disc and decreases at the location of the rings due to the increased gas surface density.
The difference near the inner boundary between the two models essentially reflects the difference in the slope of $a_{\rm max}$ in this region.
In spite of the low values of the Stokes number, the drift timescale of the grown dust is of the order of a thousand years in the inner disc, which is sufficiently short for the dust to accumulate in the pressure maxima.
{  In conclusion, the two models are in general agreement, 
however the dust+MHD model 
shows that the dead zone is more robust in terms of $\alpha_{\rm eff}$ and the dusty rings extend much farther, forming more complicated structures. }

\subsubsection{Dust evolution and Planetesimal Formation}
\label{sssec:plt}

\begin{figure*}
\includegraphics[width=0.98\textwidth]{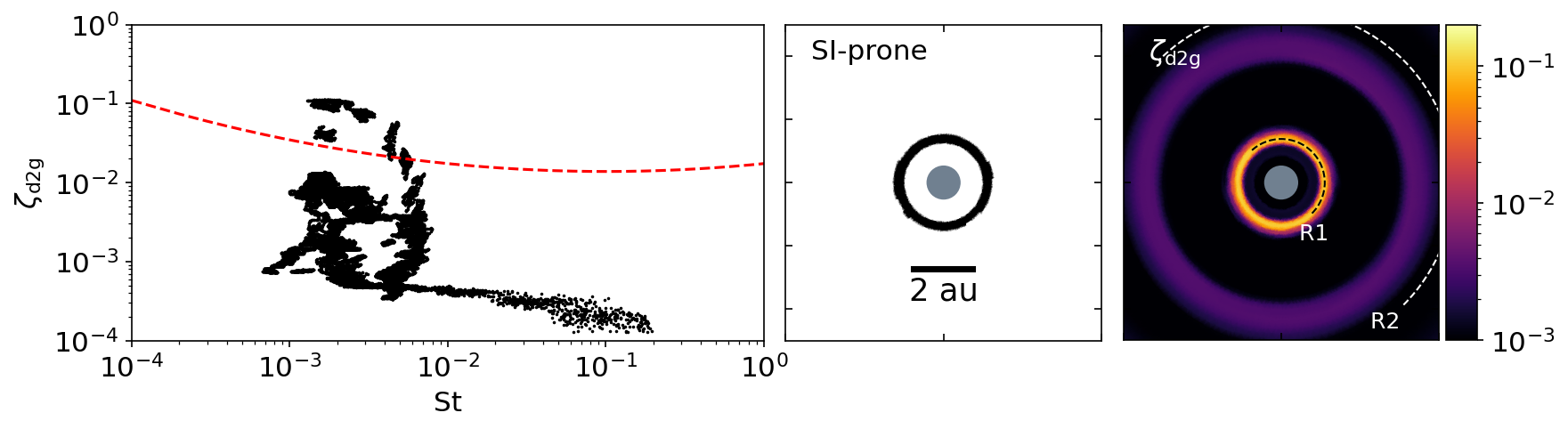}
\caption{The inner rings in \simname{model-2} are prone to streaming instability.
Left: The total dust-to-gas ratio is plotted against Stokes number for the region within the PPD at 0.35 Myr.
The red dashed line marks the critical condition for the onset of SI as per Eq. (\ref{eq:SI}). 
Right: The distribution of the region prone to SI (marked in black) as well as the total dust-to-gas ratio in the inner $10 \times 10$ au box, where the gaseous rings are marked with dashed lines.
}
\label{fig:streaming2d}
\end{figure*}

In this section, we concentrate on the dust growth, its evolution, and prospects of planetesimal formation in the fiducial dust+MHD \simname{model-2}.
While building a protoplanetary system, the growth of dust from centimeter-sized pebbles to gravitationally bound, kilometer-sized planetesimals is not well understood.
{  The increase in size of solid particles is limited due to several processes occurring in the disc, such as fragmentation and drift barrier \citep{Blum-Wurm2008,Whipple72,Weidenschilling77b}.
These barriers to dust growth may be overcome in the disc substructures wherein the pressure maxima attract dust and once the local dust-to-gas ratio reaches a sufficiently high value, streaming instability may cause the dust to spontaneously clump into planetesimals \citep{YG05,Johansen+07}. }
The SI is analogous to two-stream instability in plasmas, wherein the aerodynamically coupled dust-gas system is linearly unstable to the growth of perturbations.  
\cite{Yang+17} conducted a series of shearing-box simulations of coupled dust-gas evolution to find an empirical relation when the local conditions of the disc are suitable for the SI, i.e. when the particles spontaneously concentrate themselves into clumps.
This relation is a function of the local dust-to-gas mass ratio, $\zeta_{\rm d2g}$, and the Stokes number. 
For particles of Stokes number less than 0.1, this limit can be written as 
\begin{equation}
    {\rm log}(\zeta_{\rm d2g}) = 0.1({\rm log}\, {\rm St})^2 + 0.2 {\rm log}\,{\rm St} -1.76.
    \label{eq:SI}
\end{equation}
{ 
The canonical criterion for the onset of streaming instability is considered to be $\rho_{\rm d}/\rho_{\rm g} > 1$, where $\rho_{\rm d}$ and $\rho_{\rm g}$ are the midplane volumentric densities of dust and gas respectively \citep{YG05}.
However, recent evidence based on shearing box simulations suggests that such results of linear theory should be taken with significant caution and that the ratio of dust-to-gas surface densities similar to Eq. \ref{eq:SI} is a more robust predictor of clumping via SI \citep{Li-Youdin21, Yang+18}.
}

Figure \ref{fig:streaming2d} depicts the disc conditions for \simname{model-2} at 0.35 Myr with respect to the criterion described in Eq. (\ref{eq:SI}), when the disc is locally prone to SI and planetesimal formation.
The first panel of Figure \ref{fig:streaming2d} shows all of the points within the disc in the phase space of 
$\zeta_{\rm d2g}$ versus ${\rm St}$, while the red dashed line shows the criterion for SI.
The region of the disc that lies above this line will be susceptible for spontaneous clumping via SI.
The 2D distribution of this region is marked with black colour in the middle panel and the rightmost panel shows the dust-to-gas ratio in the same inner $10\times 10$ au region of the disc.
The enhanced value of $\zeta_{\rm d2g}$ in the first ring R1 can be observed, which corresponds to the SI-prone region in the previous two figures.
{  Note that the ratio of $\rho_{\rm d}/\rho_{\rm g}$ almost always exceeds unity in the same region of the rings and inclusion of this criterion as an additional constraint does not change the results noticeably.}
Thus, although all of the rings accumulated some amount dust, at this particular time, the condition for SI is satisfied only in the innermost ring.

\begin{figure*}
\centering
  \includegraphics[width=\textwidth]{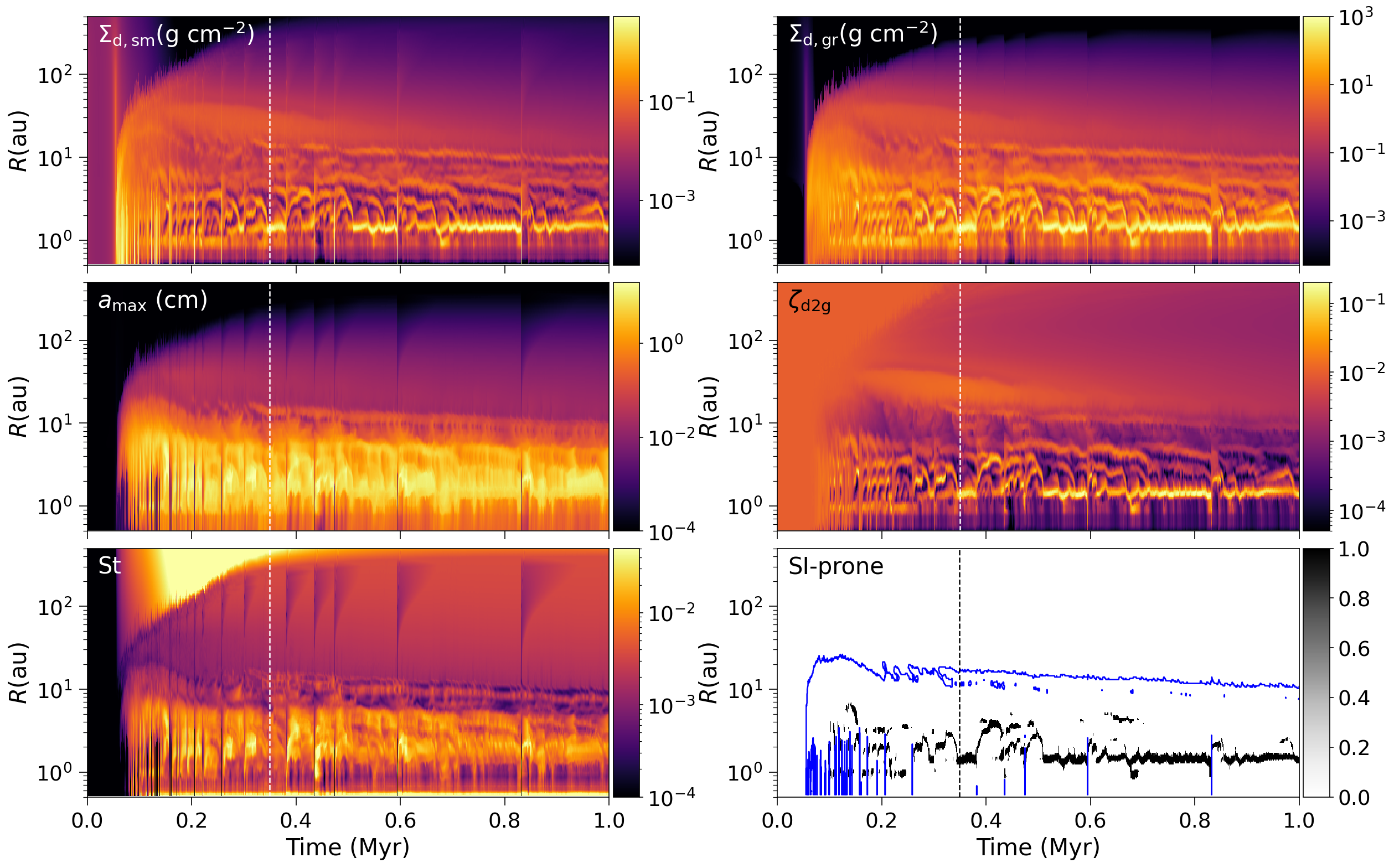}
\caption{Spacetime plots of the quantities related to the dust properties for \simname{model-2}--
surface density of small and grown dust, maximum size of dust grains, total dust-to-gas ratio, Stokes number, and the region prone to SI.
The white line in the first panel shows the extent of the disc, while the blue contour marks the dead zone.
The vertical dashed lines mark the evolution time of 0.35 Myr.
}
\label{fig:streamburst}
\end{figure*}

As we can expect from Section \ref{ssec:structure}, the dusty rings occurring in \simname{model-2} are not stationary but they evolve over time.
Figure \ref{fig:streamburst} shows the temporal evolution of the disc in terms of azimuthally averaged spacetime diagram of quantities related to its dust properties.
The first row of Figure \ref{fig:streamburst} shows the evolution for the small and grown dust surface densities. 
The grown dust approximately follows the distribution of the small dust, although with some noticeable differences.
At very early times, most of the dust mass is in $\Sigma_{\rm d,sm}$, although soon after disc formation the dust rapidly grows and $\Sigma_{\rm d,gr}$ dominates $\Sigma_{\rm d,sm}$ by several orders of magnitude.
Since the small dust continually gets converted to grown dust in the dead zone because of favourable conditions, $\Sigma_{\rm d,sm}$ shows bands of depletion in the spacetime plot.
The second row of Figure \ref{fig:streamburst} shows the maximum size of the dust particles and the total dust-to-gas mass ratio, including both small and grown dust.
The dust exceeds typical pebble size of 0.5 cm and grows up to a few centimeters across. Such grown dust particles are found not only in the rings but in most of the inner regions of the dead zone.
The dust-to-gas ratio and Stokes number in the last row are the quantities that determine the region that is prone to SI.
In the inner few au of the disc, $\zeta_{\rm d2g}$ consistently exceeds 0.1, which is a significant enhancement over the initial ratio of 1\%.
Since the grown dust dominates the total dust mass, $\zeta_{\rm d2g}$ mainly reflects its enhancement with respect to the distribution of gas.
The Stokes number is calculated at $a_{\rm max}$.
The Stokes number increases to a large value outside the extent of the disc due to a large decrease in the gas volume density, although note that typical assumptions for calculating St also break down outside of a Keplerian disc.
The Stokes number is high in the dead zone, although it remains below 0.1 and does not trace the distribution of the dust or the gas.
The last panel of Figure \ref{fig:streamburst} marks the region of the disc prone to SI in black.
As described by Eq. (\ref{eq:SI}), the criterion for SI is based on the local values of $\zeta_{\rm d2g}$ and ${\rm St}$.
The inner regions of the dead zone are prone to SI throughout the 1 Myr evolution of the PPD.
However, typically only the innermost dusty ring is consistently conducive to SI.
At early times and outside of the innermost ring, relatively transient regions of SI are occurring.

The blue line in the last panel of Figure \ref{fig:streamburst} traces the extent of the dead zone.
The innermost few au region of the disc is occasionally punctuated by narrow spikes in this contour, which were also coincident with sharp discontinuities in the other spacetime diagrams.
{  These events correspond to the episodic accretion caused by MRI-type outbursts, elaborated in Section \ref{sssec:MRIburst}. }
At early times, the disc is also susceptible to gravitational clumps and spirals, which can cause similarly luminous bursts due to clump accretion and rapid stochastic variation in the accretion rate \citep{VB15,Vorobyov-Elbakyan18}. 
Although not all such events are captured in the blue dead zone contour due to the sampling of the output data, the SI-prone region seems to be punctuated by the MRI outbursts.

\subsubsection{Dust evolution with MRI Outbursts}
\label{sssec:MRIburst}

\begin{figure*}
\begin{tabular}{l}
\vspace{-0.21cm}\includegraphics[width=\textwidth]{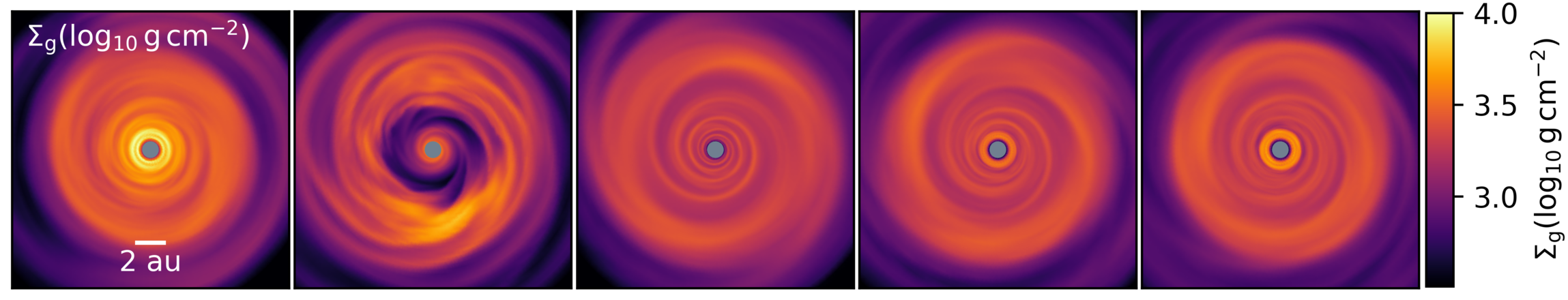} \\ 
\vspace{-0.21cm}\includegraphics[width=\textwidth]{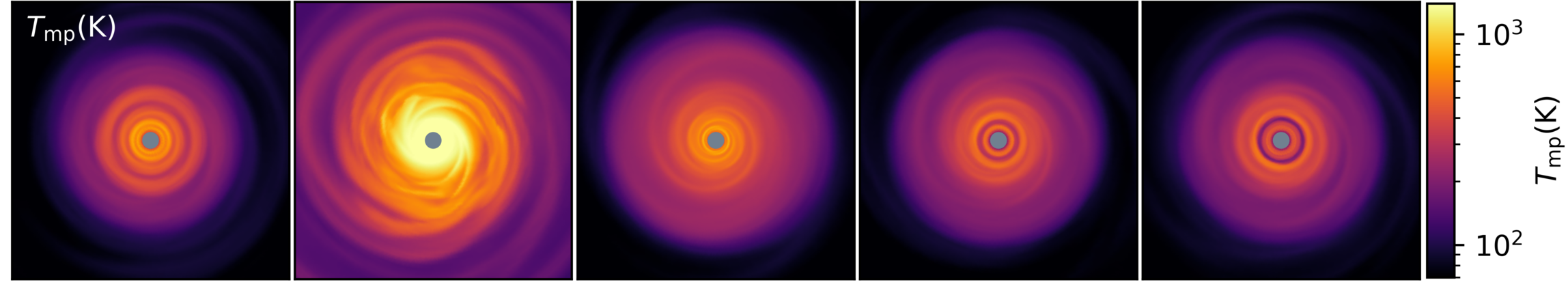} \\
\vspace{-0.21cm}\includegraphics[width=1.01\textwidth]{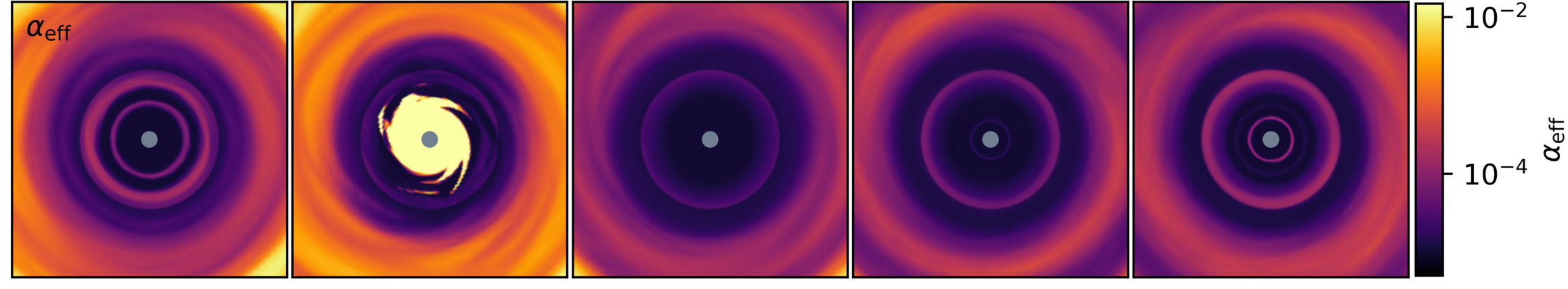} \\
\vspace{-0.21cm}\includegraphics[width=\textwidth]{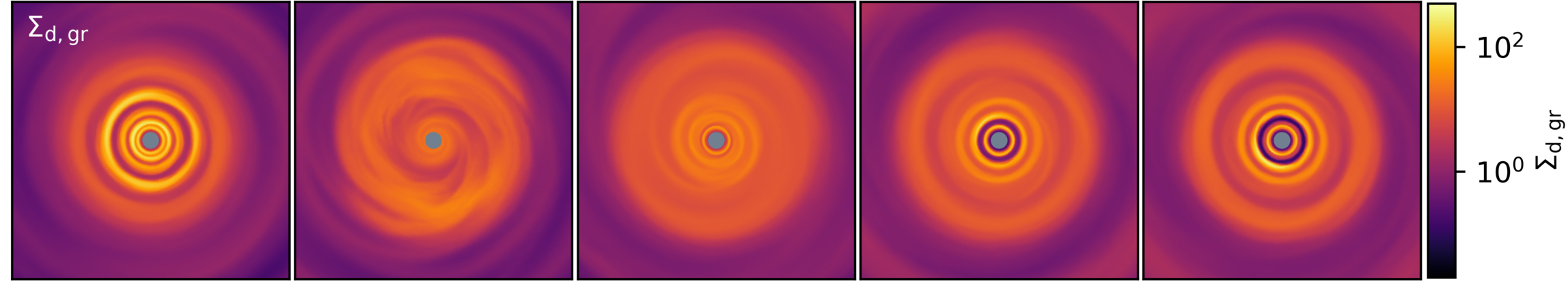}\\
\vspace{-0.28cm}\includegraphics[width=1.01\textwidth]{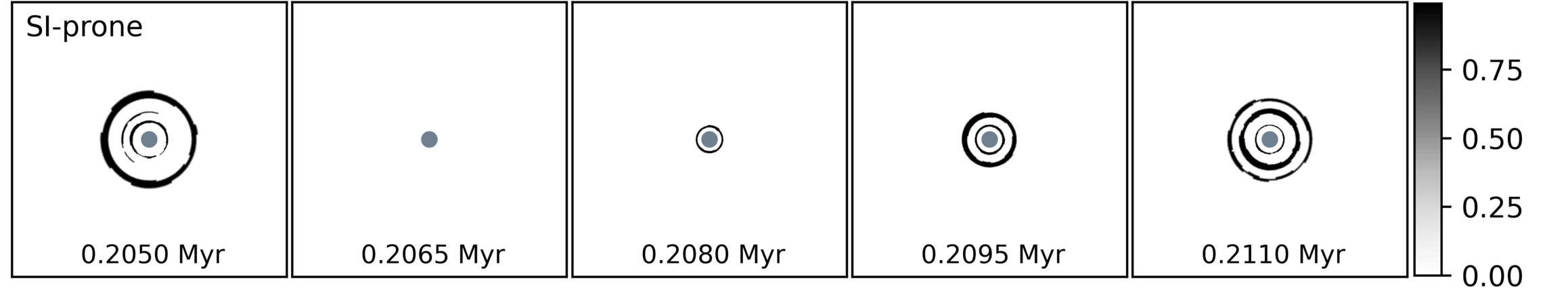}\\
\hspace{-0.46cm}\includegraphics[width=0.915\textwidth]{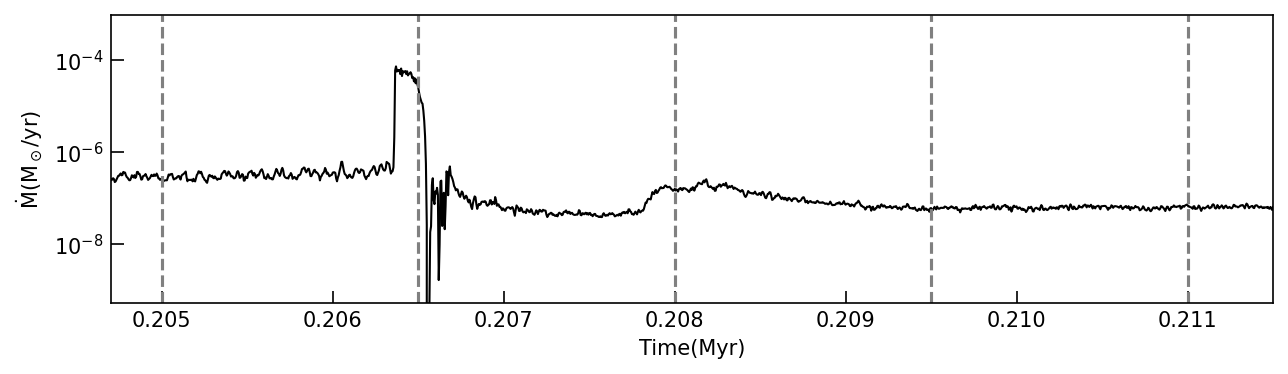}
\end{tabular}
\caption{The temporal evolution of the distribution of the gas surface density, midplane temperature, $\alpha_{\rm eff}$, surface density of the grown dust, and the region prone to SI in the inner $10\times10$ au box of \simname{model-2}, showing the progression of a typical outburst at $\approx 0.2$ Myr. The bottom panel shows the mass accretion rate on to the central star. The vertical dashed lines mark the time of the aforementioned distributions.} 
\label{fig:burst2d}
\end{figure*}

{ 
The episodic accretion perturbs the inner disc significantly and its effects on the dusty rings are worth investigating in more detail.
The inner disc undergoes MRI bursts similar to observed FU Orionis events \citep[e.g. elaborated in][]{Kadam+20,VMHD20,Zhu+10a}.
In summary, the small amount of viscosity in the dead zone eventually raises the local temperature above $\sim 1000$ K. At this temperature, the alkali metals are ionized, increasing the ionization fraction exponentially and making the disc suddenly MRI-active.
The ionization front travels outwards in a snowplow fashion and the accumulated material in this area is accreted on to the star, resulting in a powerful outburst.}

Figure \ref{fig:burst2d} focuses on the evolution of the inner disc structure across a typical MRI burst occurring at about 0.2 Myr.
The rows show the 2D distributions of some of the key quantities separated in time by 1500 years.
The lightcurve is plotted in the bottom panel in terms of the average accretion rate on to the central star. 
The vertical dashed lines mark the time corresponding to the 2D slices shown above.
During the outburst, the accretion rate rapidly increases from a few times $10^{-7} \Mdot$ to nearly $10^{-4} \Mdot$.
The outburst duration is about 200 yr and it shows a highly asymmetrical profile with a sharp rise and slow decline, typical of observed FU Orionis light curves \citep[e.g. V1057 Cyg,][]{Audard+14}.
The first two time slices are captured before and during the outburst, respectively, while the next three correspond to post-outburst.
Consider the first row of Figure \ref{fig:burst2d}, showing the distribution of $\Sigma_{\rm g}$.
The gas surface density is greater in the inner regions before the outburst, while a cavity spanning over 2 au in radius can be observed past the peak of the burst.
This gas has already accreted on to the star, fueling the current accretion rate.
The next panel after the eruption shows that the cavity is soon filled with material flowing from the outer disc, although $\Sigma_{\rm g}$ remains approximately constant thereafter.
The outburst is clearly seen in the second and third rows, which show midplane temperature and $\alpha_{\rm eff}$, respectively.
The temperature in the inner disc exceeds $1300$ K, and due to the ionization of alkali metals, the disc becomes fully MRI active.
Note that this temperature threshold for the ionization of alkali metals was set explicitly in the gas only model as well as similar studies of episodic accretion \cite[e.g.][]{Bae+14}.
In the case of dust+MRI simulations, including \simname{model-2}, the degree of thermal ionization arises from the low ionization potential of potassium and is added to the ionization fraction calculated from Eq. (\ref{eq:ion}).
Both $T_{\rm mp}$ and $\alpha_{\rm eff}$ show only marginal changes after the outburst.
The fourth row shows the surface density of the grown dust.
The dusty rings are clearly seen to have formed in the first time frame, all of which get disrupted during the burst.
Over the next few thousand years, the small dust is replenished from the outer disc and it starts to grow into rings of grown dust in the inner disc.
The last row of the 2D distributions shows the region prone to SI.
Multiple rings and arches are suitable for developing SI in the initial frame. 
The conditions become entirely unsuitable for SI during the outburst, as accumulated grown dust is accreted on to the star along with the gas component.
The SI-prone region starts reappearing soon and within a few thousand years after the burst, stable concentric rings are again developed.
These SI-prone rings do not necessarily trace $\Sigma_{\rm d,gr}$, but follow the suitable conditions in terms of the dust-to-gas ratio and Stokes number. 
The disruption of the SI-prone region and accretion of dust during outbursts implies that the episodic accretion may have a direct and significant impact on the process of planetesimal and planet formation.

\begin{figure}
\includegraphics[width=\columnwidth]{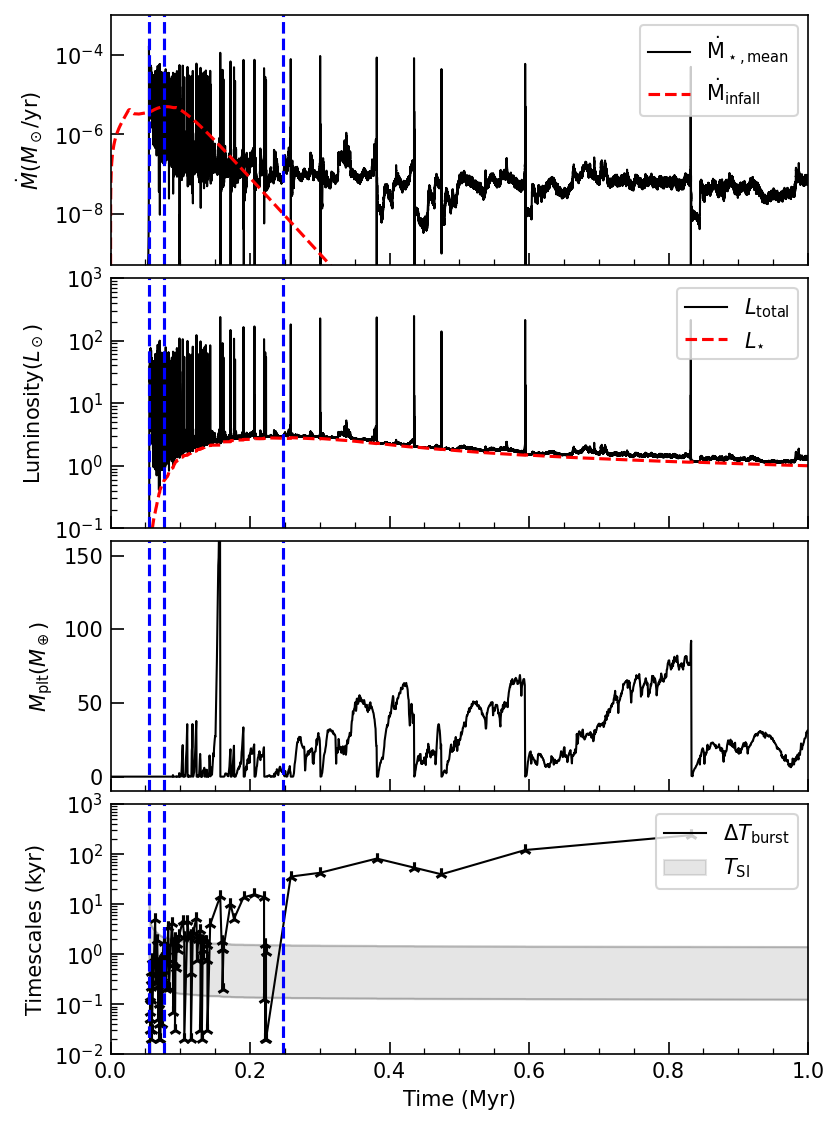}
\caption{Effects of episodic accretion on the evolution of the SI prone region over time for \simname{model-2}.
The top panel shows mean accretion rate on to the star (black) and average cloud core infall rate (red, dashed). The second panel shows total (black) and stellar (red, dashed) luminosity. The third panel shows the mass of the grown dust that is prone to SI over the entire disc. The last panel compares the time between significant outbursts to the approximate growth timescale of SI depicted with the grey band. The three vertical dashed lines mark the time of star formation and end of Class I and Class II stages, respectively.
}
\label{fig:SItime}
\end{figure}

We demonstrate the long-term effects of episodic accretion on planetesimal formation with the help of Figure \ref{fig:SItime}.
The vertical lines mark the time of formation of the central protostar, the end of Class 0 stage, and the end of Class I stage, respectively.
The Class 0/I boundary is defined to be the point at which the gas mass of the star and disk system exceed half of that of the initial cloud core.  
The Class I/II boundary is considered as the time when the envelope accretion rate falls below $10^{-8}\Mdot$.
The first panel shows the mass accretion rate on to the star as well as the cloud mass infall rate. 
The second panel shows the stellar luminosity, which forms a baseline to which the accretion luminosity is added to obtain total luminosity of the system.
In these plots, the episodic accretion of the system can be clearly seen in terms of intermittent outbursts.
The accretion rate increases to nearly $10^{-4} \Mdot$ and the total luminosity increases to about 100 ${\rm{L_\odot}}$ during a typical outburst. 
In the early stages when the cloud infall rate is comparable to the accretion rate and the disc mass is relatively large as compared to the stellar mass, the disc is prone to gravitational instability.
This results in spirals and clumps and associated variation in accretion rate is reflected in the forest of frequent luminosity variations during the embedded phase.
The rapid fluctuations in the accretion rate at this time can also correspond to the excursion of the inner fully MRI-active region across the computational boundary \citep{Kadam+21}.

The third panel in Figure \ref{fig:SItime} shows the instantaneous value of the total mass of grown dust in excess of that required for SI in the SI-prone region,
\begin{equation}
    M_{\rm plt}= \sum_{\rm SI-prone} \frac{\zeta_{\rm d2g} - \zeta_{\rm d2g, crit}}{\zeta_{\rm d2g}} M_{\rm d,gr}, 
\end{equation}
where $\zeta_{\rm d2g, crit}$ is the critical threshold of dust-to-gas ratio for SI given by Eq. \ref{eq:SI} and $M_{\rm d,gr}$ is the mass of the grown dust in that cell.
Note that this is the maximum mass in the form of grown dust that is available for planetesimal formation throughout the disk at any given time.
The last panel shows the duration between the successive bursts of significant magnitude ($\Delta T_{\rm burst}$).
A threshold of $50\Lsun$ is chosen to define bursts in this study, as all outbursts at later times as well as initial luminosity variations are comfortably above this value.
Dust-gas coupled hydrodynamic simulations in the shearing box approximation suggest that the timescale of growth for SI is of the order of 100 times the local orbital period \citep{Simon+16}.
Since the SI-prone region is limited to the dead zone and extends between the innermost 1 and 10 au of the disc, the grey region covers the SI growth timescale between these two radii, i.e. between 100 times the orbital period at these two radii at a given time.
The disc contains a significant amount of mass in the dusty rings in terms of grown dust that is available for planetesimal and hence eventual planet formation. 
At 0.8 Myr, up to $70\Mearth$ of grown dust is prone to the SI, which is comparable to the total mass of solids in our solar system \citep{Weidenschilling77a}.  
The disc environment becomes favourable for planetesimal formation very early, during the embedded Class I stage.
The dust mass available for planetesimal formation is destroyed during the outbursts, as reflected in the abrupt decline in $M_{\rm plt}$ coincident with the accretion bursts.
In the initial stages, the outbursts are much more frequent and the time duration between bursts is comparable with the growth timescale of SI.  
As the disc evolves and the envelope mass infall rate decreases, the bursts become increasingly separated in time and the period between bursts exceeds the SI growth timescale by two orders of magnitude.
This implies that planetesimals would indeed have sufficient time to form within the local environment of the dusty rings.
Once planetesimals are formed, they are essentially decoupled from the gas because of their large Stokes number value and would not be affected by the gas dynamics of the accretion bursts.
Thus, the process of planetesimal formation may be a tug of war between the dusty rings that promote SI and episodic outbursts that can destroy these reservoirs of dust.

\subsection{Effects of Magnetic Field and Mass }
\label{ssec:param}

\begin{figure*}
\centering
  \includegraphics[width=0.99\textwidth]{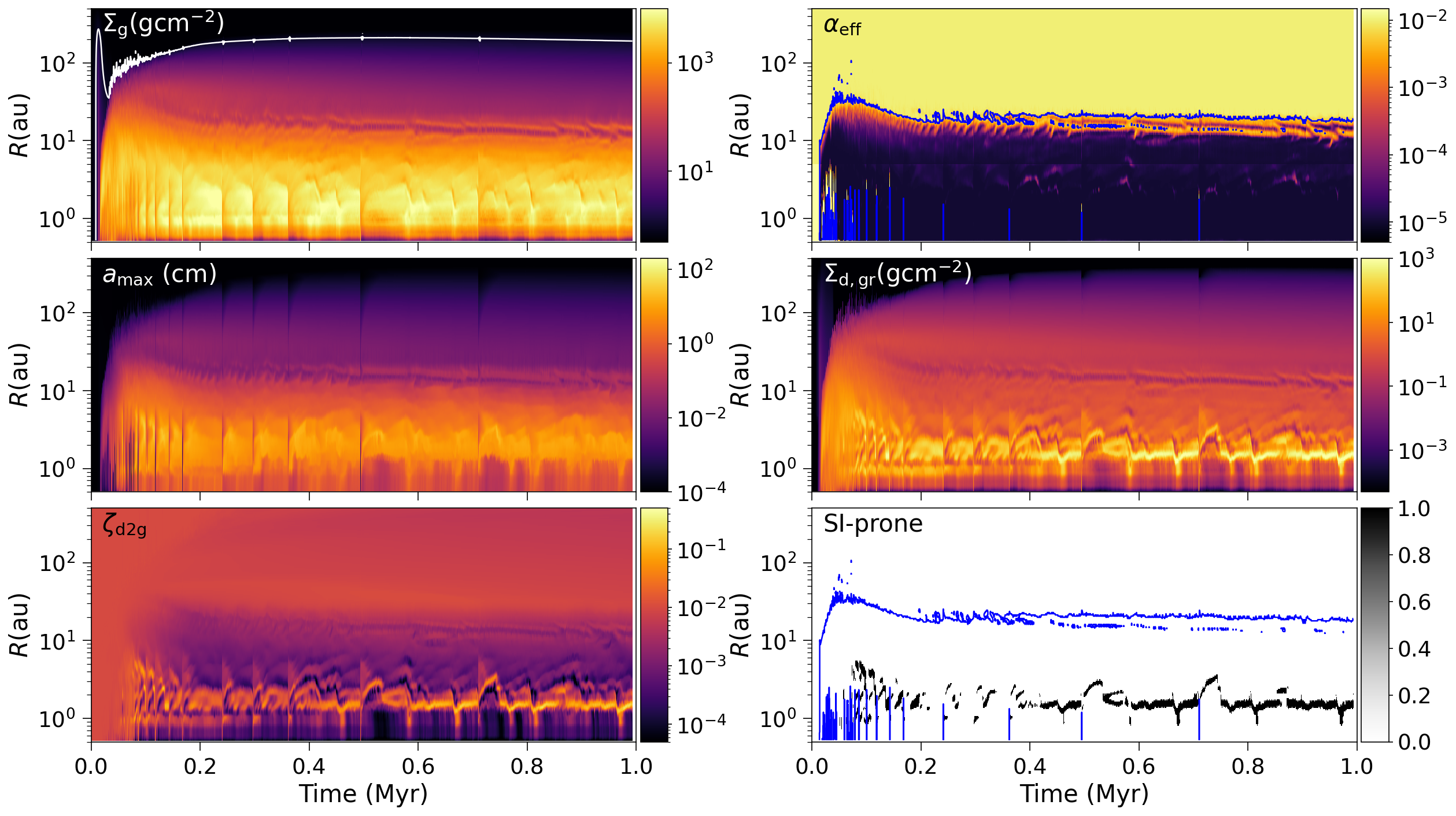}
\caption{Spacetime plots of the quantities related to the dust properties for high $\lambda$ \simname{model-2L}--gas surface density,$\alpha_{\rm eff}$, $a_{\rm max}$, grown dust surface density, total dust-to-gas ratio, and region prone to SI.
The white curve in the first panel shows the outer boundary of the disc, while the blue curves mark the extent of the dead zone.
}
\label{fig:model2L}
\end{figure*}

\begin{figure}
\includegraphics[width=0.93\columnwidth]{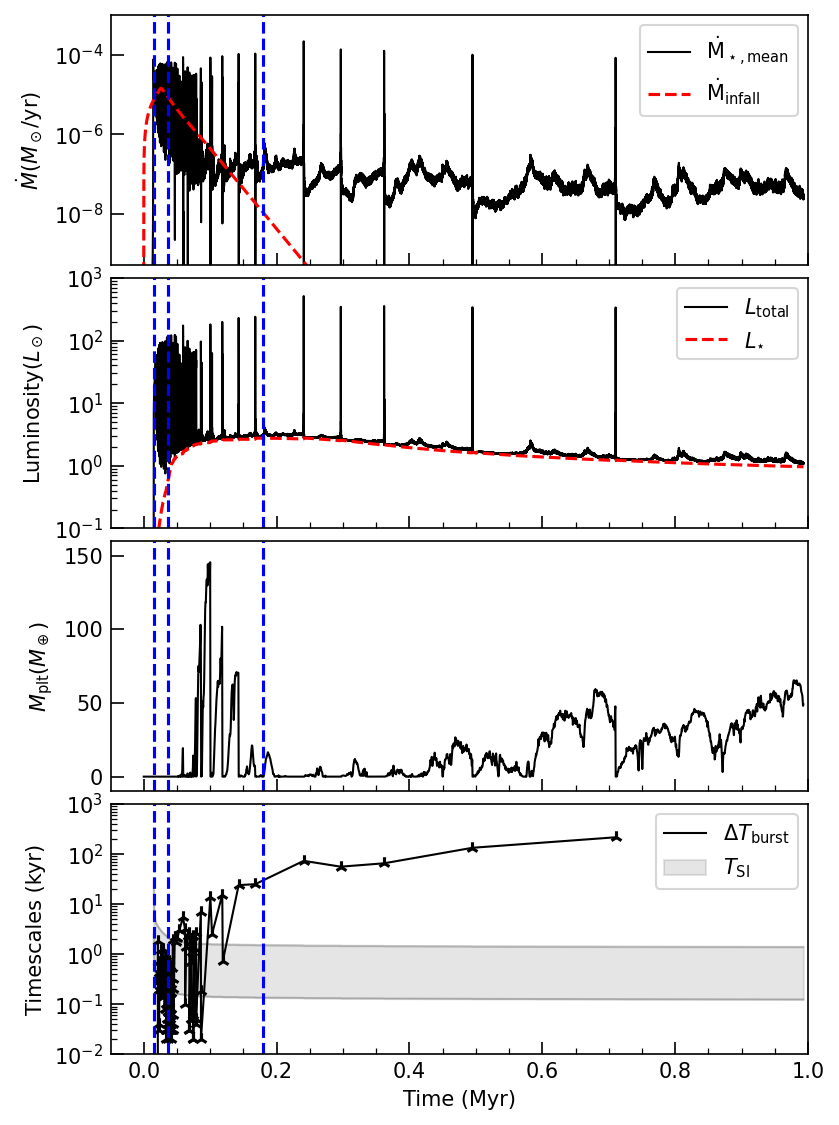}
\caption{Similar to Fig.~\ref{fig:SItime}, but for high $\lambda$ \simname{model-2L}.}
\label{fig:model2LSI}
\end{figure}

In this section we present the results of a limited parameter study with a focus on the evolution of the dusty rings and the prospects of planetesimal formation.
Observations suggest that the mtf ratio $\lambda$ is typically $\gtrsim 1$ \citep{crutcher2012,myers2021,liu2022} for molecular cloud cores, so we adopt a value of 2 in the fiducial \simname{model-2} and a higher value of 10 for \simname{model-2L} to account for some increase of $\lambda$ in the protostellar disc phase.
Figure \ref{fig:model2L} shows the spacetime diagrams of some of the selected quantities that are relevant to understanding the dusty rings in \simname{model-2L}. 
The first row shows the gas surface density and $\alpha_{\rm eff}$.
With an increased $\lambda$, the magnetic support to the cloud core is reduced. 
As a result, the disc forms earlier at 0.015 Myr, as compared to 0.06 Myr for \simname{model-2}.
The general evolution of $\Sigma_{\rm g}$ and the dead zone is similar to that of the fiducial model.
However, note that the $\alpha_{\rm eff}$ values in the dead zone are marginally lower; this is because the $\Sigma_{\rm MRI}$ in the model is proportional to $B_z$ (Eq. \ref{eq:DZ}), which is lower in the case of \simname{model-2L}.
The second row of Figure \ref{fig:model2L} shows $a_{\rm max}$ and grown dust surface density for \simname{model-2L}.
Similar to \simname{model-2}, the innermost ring in this model shows consistently large concentrations of dust.
The dusty rings in the inner regions can be seen in both in $\Sigma_{\rm d, gr}$ and $\zeta_{\rm d2g}$.
The overall structure and evolution of the inner region is very similar to fiducial \simname{model-2}.
The dust-to-gas ratio is as large as 0.1 in the innermost ring, while this region was also consistently conducive to SI.
However, this SI-prone region is noticeably more disconnected in the early evolution of the disc, before approximately 0.4 Myr.
With a greater $\lambda$ value, the dead zone is more robust, since $\Sigma_{\rm crit}$ quadratically depends on the magnetic field.
In comparison with \simname{model-2}, a lower $\Sigma_{\rm crit}$ implies a greater bottleneck for angular momentum transport, which results in marginally larger concentration of gas in this region.
This leads to a decrease in both the dust-to-gas ratio and Stokes number in such a way that the disc is less susceptible to SI.

Figure \ref{fig:model2LSI}, which is similar to Figure \ref{fig:SItime}, shows the episodic accretion in \simname{model-2L} with respect to the prospects of planetesimal formation.
Note that the abscissa starts at a negative value in this plot to make some of the features at early times more clear.
The faster evolution of the system can be observed in the first panel of $\dot{M}$, where the duration of cloud infall as well as the initial forest of rapid bursts is shorter than \simname{model-2}.
The overall evolution of the total luminosity, including episodic outbursts of large magnitude is similar to the fiducial model.
The mass available for planetesimal formation peaks several times during the Class I stage; although the time between outbursts, as seen in the last panel, is short and comparable to SI growth timescale at this time.
For \simname{model-2L}, the dust mass available for planetesimal formation is not significantly large before about 0.4 Myr, presumably because of the marginal increase in the surface density of the accumulated gas.
The conditions become favourable subsequently during the Class II stage and up to 50 $M_{\rm \oplus}$ of grown dust mass is available for planetesimal formation near the end.
The dust remains in the disc for sufficient time for SI to set in before it is destroyed by the bursts, as $\Delta T_{\rm burst}$ remains comfortably above the SI growth timescale.
In conclusion, the increased $\lambda$ marginally affects the dusty rings such that $\zeta_{\rm d2g}$ and St are relatively low in the early evolution, which possibly implies late or decreased planetesimal formation.

\begin{figure*}
\centering
  \includegraphics[width=0.99\textwidth]{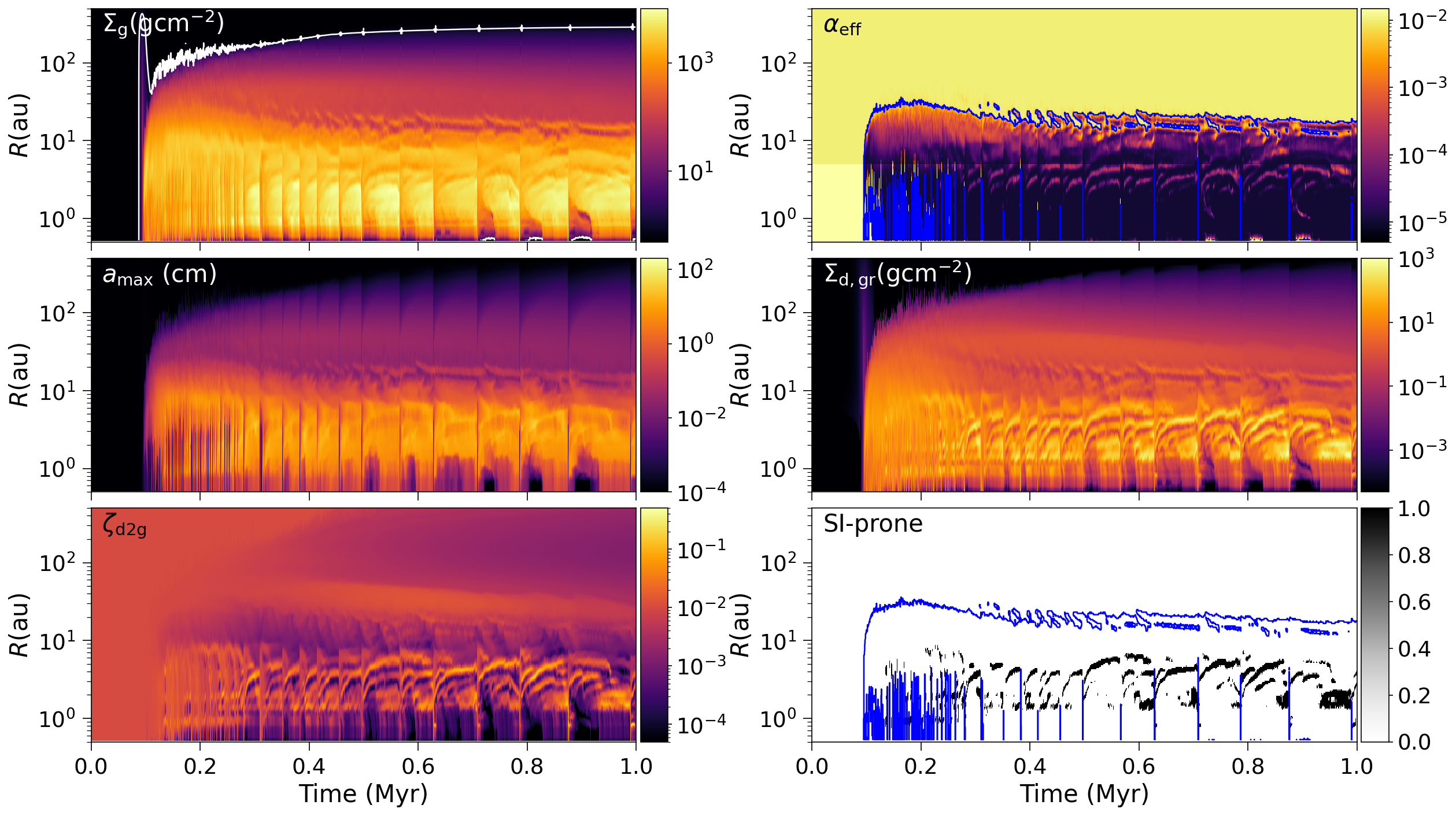}
\caption{Similar to Fig.~\ref{fig:model2L}, but for high mass \simname{model-1}.}
\label{fig:model1}
\end{figure*}

\begin{figure}
\includegraphics[width=0.94\columnwidth]{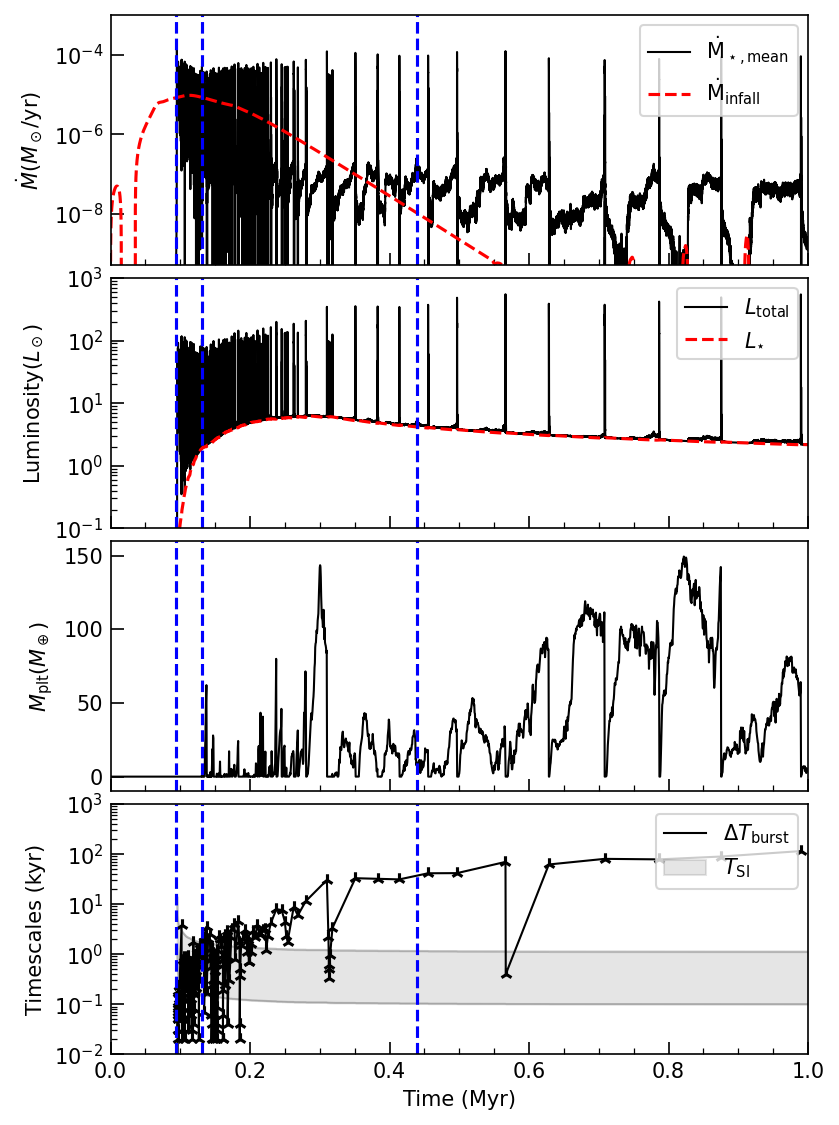}
\caption{Similar to Fig.~\ref{fig:SItime}, but for high mass \simname{model-1}.}
\label{fig:model1SI}
\end{figure}

We now consider the effects of varying stellar mass on the properties of the dusty rings. As described in Table \ref{table:sims}, \simname{model-1} has a mass of $1.45 \Msun$, but is otherwise identical to the fiducial \simname{model-2}. 
Figure \ref{fig:model1} shows the spacetime diagrams of this high mass \simname{model-1}.
The overall structure and evolution of the disc is similar to the fiducial model.
Intricate rings formed in the inner regions of the disc, which can be seen in the grown dust surface density.
The dust-to-gas mass ratio is enhanced in these dusty rings and the local conditions are suitable to develop SI.
The rings are periodically disrupted by outbursting events observed in the contours of the dead zone as well as discontinuities in the spacetime diagrams.
Despite these similarities, \simname{model-1} shows some key differences in comparison with \simname{model-2}.
The first difference is that the outbursts are more frequent.
The larger initial mass of the cloud core generally results in a relatively more massive disc.
Stronger gravitational instability associated with a more massive discs can replenish the inner disc quickly after an outburst, rendering the outbursts more frequent in this high mass model.
The second major difference is that the SI-prone region forms at a noticeably larger radial distance from the central star.
As the grown dust migrates inward in the high mass disc, the dust-to-gas ratio exceeds the threshold for SI at a larger radial distance.
However, the gas surface density keeps increasing inward and in the innermost regions this results in unfavourable conditions for SI due to relatively low values of the Stokes number.

Figure \ref{fig:model1SI} shows the mass accretion history as well as the mass available for planetesimal formation for \simname{model-1}.
In comparison with the fiducial \simname{model-2}, the envelope infall lasts for a longer time, resulting in a longer embedded phase.
From the plots of accretion rate as well as total luminosity, the increased outburst frequency is clearly observed in \simname{model-1}.
The maximum in $M_{\rm plt}$ increases significantly to about $150\Mearth$ at about 0.8 Myr, which is nearly twice that found in \simname{model-2}.
Although the rings are located at a larger distance for \simname{model-1}, they are still perturbed by the eruptions, as $M_{\rm plt}$ displays concurrent discontinuities.
The last panel shows that despite the higher outburst frequency, the SI growth timescale is considerably shorter than the time period between outbursts after about 0.2 Myr.
At this time, the system is still in the early Class I stage.
Thus, we conclude that the higher mass system exhibits episodic accretion that is more frequent, while the dusty rings are formed at marginally larger radius and contain more mass for planetesimal formation.
However, the overall disc behaviour in terms of outbursts and their effects on dusty rings as well as SI growth during quiescence remains unchanged.

\begin{figure*}
\centering
  \includegraphics[width=0.99\textwidth]{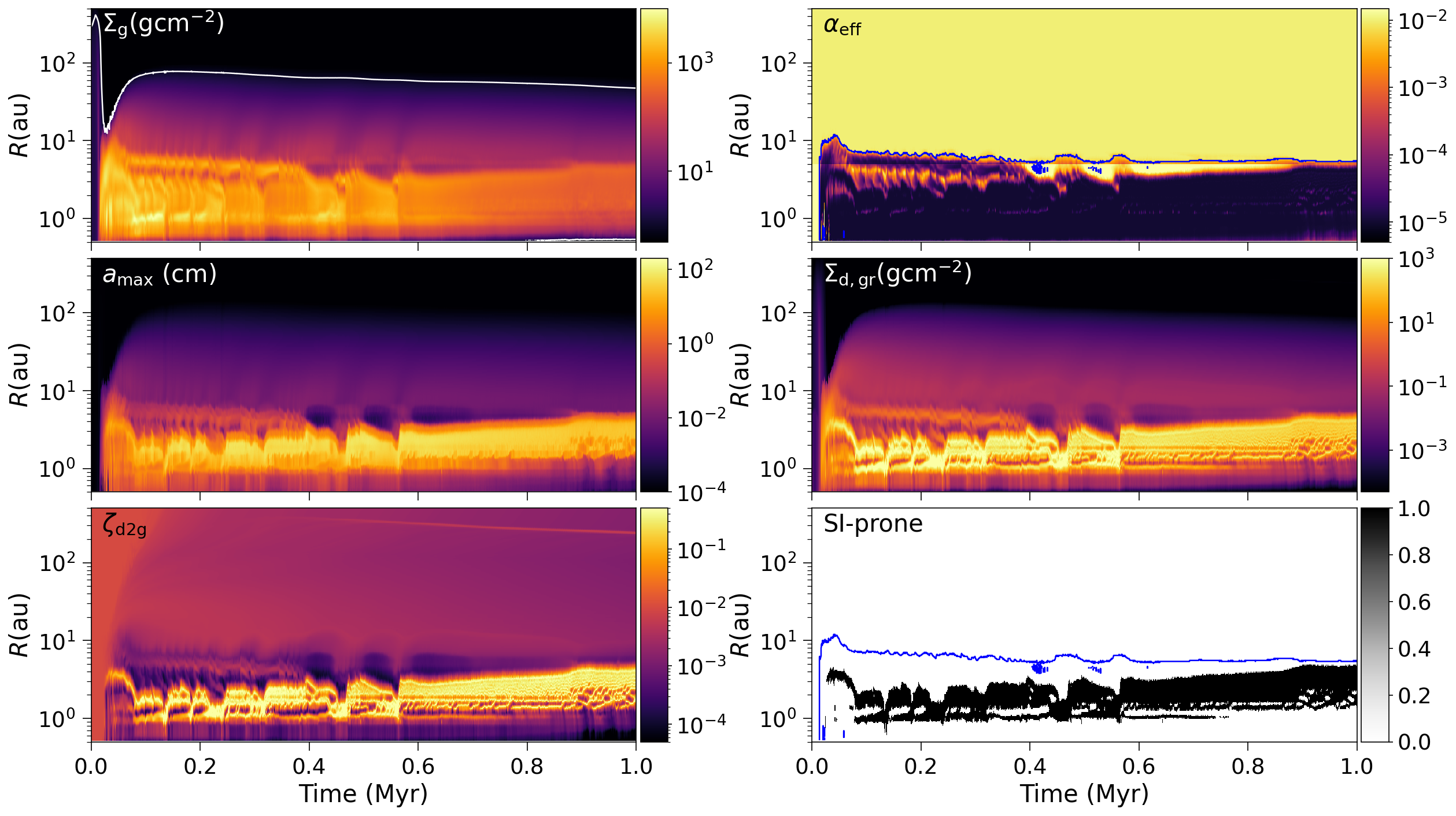}
\caption{Similar to Fig.~\ref{fig:model2L}, but for low mass \simname{model-3}.}
\label{fig:model3}
\end{figure*}

\begin{figure}
\includegraphics[width=0.93\columnwidth]{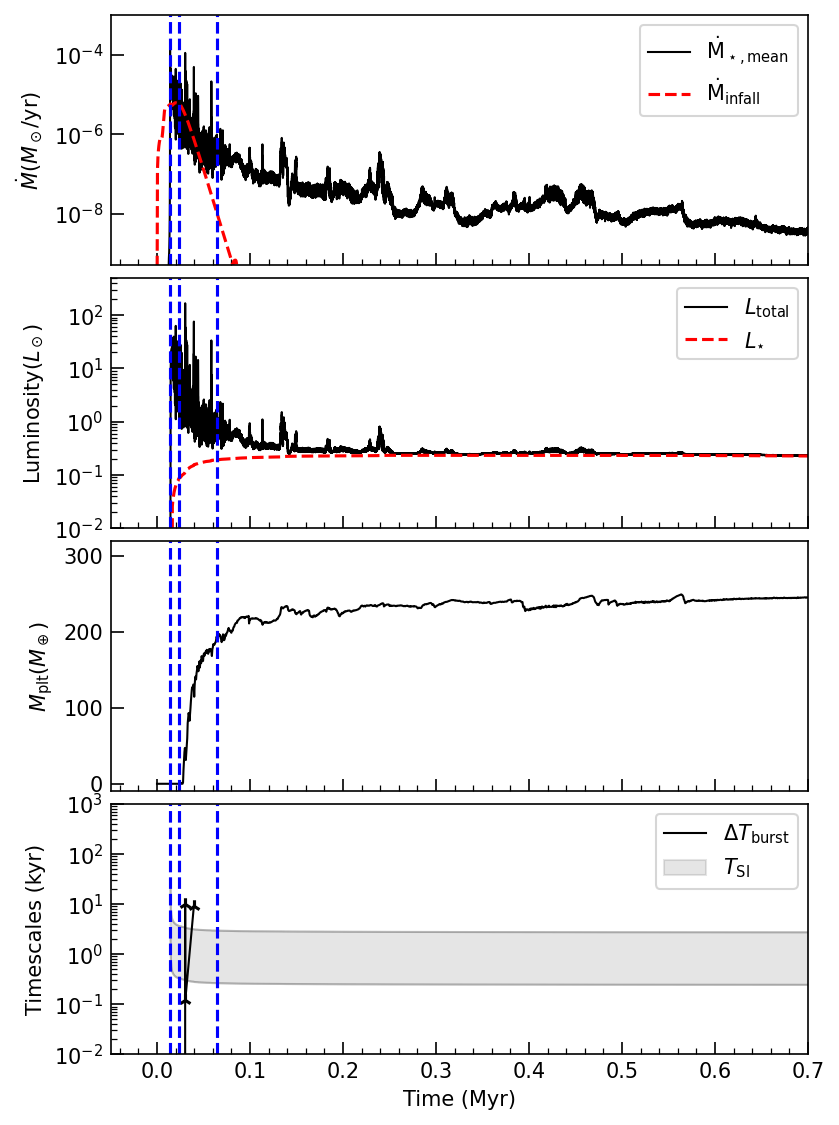}
\caption{Similar to Fig.~\ref{fig:SItime}, but for low mass \simname{model-3}.}
\label{fig:model3SI}
\end{figure}

The final simulation in our parameter-space is the low mass \simname{model-3}, with a total initial gas mass of 0.21$\Msun$. 
Consider Figure \ref{fig:model3}, where the disc forms much earlier due to the shorter collapse time.
The outer extent of the dead zone as well as the disc size is considerably smaller as compared to the fiducial \simname{model-2}. 
However, the dead zone formed is similarly robust in terms of low values of $\alpha_{\rm eff}$.
The inner disc develops substructures in terms of gas and dust, although the dusty rings in $\Sigma_{\rm d,gr}$ are wider than in previous simulations. 
The wider rings essentially reflect the dust-gas dynamics of a disc with low gas surface density. 
The lower gas surface density results in efficient cooling of the disc, which leads to much lower midplane temperatures than any of the previous models at this time.
The decreased temperature has significant effects on the dust evolution.
The maximum size of the dust particles, $a_{\rm max}$, increases with lowering of temperature and consequently, a larger fraction of the available dust is converted into the grown component.
This results in a larger dust-to-gas ratio as seen in Figure \ref{fig:model3}.
The Stokes number is also inversely proportional to the temperature; as ${\rm St}$ increases, the gradiental drift velocity of the grown dust also increases, making its accumulation easier in the  pressure traps.
The increase in both $\zeta_{\rm d2g}$ and St ultimately results in a more radially extended SI-prone region.

The accretion history and plots relevant for planetesimal formation for \simname{model-3} are shown in Figure \ref{fig:model3SI}. 
The shorter duration of the cloud core infall can be seen in the first panel, along with the shortened embedded Class 0 and I stages.
The low mass model shows a key and qualitative difference in terms of episodic accretion--the outbursts are nearly absent, especially at later times during the Class II stage.
This scarcity of bursts is consistent with the results presented earlier with gas-only simulations in \cite{Kadam+19}.
The limited mass reservoir in \simname{model-3} results in a relatively lower gas surface density. 
Such a disc typically cannot trigger MRI type outbursts due to efficient cooling caused by the lower optical depth.
Although three distinct eruptions are seen at very early times, these do not perturb the disc significantly, as the $M_{\rm plt}$ curve does not show corresponding discontinuities. 
Due to the scarcity of outbursts, the grown dust continues to stay in the disc, while because of the wider rings, the mass available for planetesimal formation is as large as about $220\Mearth$.
Since the outbursts cease at early times, most of the grown dust prone to SI
should eventually get converted into planetesimals.
Thus, we find that the environment in the disc surrounding this low mass star is particularly conducive for planetesimal formation via SI due to both dust-gas dynamics and rarity of outbursts.

\section{Discussion}
\label{sec:discussions}
The episodic accretion in PPDs has been a subject of several theoretical studies, however their long-term ($\sim 1$ Myr) evolution is tackled by only a handful of investigations \citep{VB06,Bae+14,VMHD20,Kadam+20}.
Our approach to this problem using the FEOSAD code is unique and it has several advantages.
The simulations start with the core collapse phase of a magnetized starless molecular cloud core, so that the ultimate disc formation is self-consistent. 
The system evolves into the embedded phase of star formation, during which the central star and its disc are formed.
This eliminates the typical assumption of a minimum mass solar nebula \citep[MMSN;][]{Adams10} as an initial condition, which may not be suitable for exoplanet systems in general.
In order to capture the intricate structures formed in the inner disc region, the inner boundary of the computational domain is placed at 0.52 au, with special inflow-outflow boundary conditions.
The small inner boundary allows the disc to undergo MRI outbursts, while the dynamics of gravitational instability are also captured.
The evolution of magnetic fields is taken into account in the flux freezing limit, while the adaptive $\alpha$-parameter is derived from explicit calculations of the ionization fraction.
The coevolution of coupled dust is also taken into account, including its back-reaction on gas and dust growth.
Thus, the insights obtained from such a framework cannot be gained by 1D simulations, or 3D simulations that are limited to a short duration of time and spanning a partial region of the disc (e.g. extending a few tens of au of a PPD or using a shearing box). Our simulations also avoid starting with MMSN initial conditions and are able to include increasingly complex physics.

Our investigation paints an intriguing picture of the process of planet formation.
The planetesimals are produced in the primordial dusty rings formed in the innermost parts of a protoplanetary disc.
The local conditions in the rings become conducive for planetesimal formation very early during the embedded Class I stage of star formation.
As the disc evolves, the available mass for planetesimals increases and is typically sufficient to form planetary systems similar to our solar system.
The episodic accretion has significant impact on the inner disc and hence the planetesimal forming regions in the rings.
During the powerful MRI outbursts, the accumulated dust, as well as gas, in the rings gets disrupted and accretes on to the star.
The time between the bursts becomes longer than the growth timescale for SI typically before the beginning of Class II stage.
This implies that the conditions in a PPD become well suited for planetesimal and hence planet formation as early as the Class I stage.
However, note that once formed, the planetesimals should decouple from the gas dynamics and will not be affected by the outbursts.
The further evolution of the system may result in inside-out planet formation, wherein planets in the inner disc are formed in situ by opening up sequential gaps \citep{CT14}.
Growing observational evidence suggests that planet formation takes place before the Class II phase of a PPD \citep{Andrews-Williams07,Tychoniec+20}.
The planetesimal formation in the dusty rings during the embedded phase is consistent with these findings.
Exoplanet surveys indicate that the low mass stars have a higher planet occurrence rate of Neptunes and super-Earths, especially in multiple systems \citep{Cassan+12,Hatzes16}.
As PPD masses decrease for lower mass stars, this fact is challenging to explain with conventional theories of planet formation, including in-situ formation and planetary migration \citep{Andrews+13,Mulders+15,Pascucci+16,Vorobyov11}.
Our result of more favourable conditions in low mass discs due to dust-gas dynamics and paucity of episodic outbursts naturally explains these observations.
This proposed scenario for planet formation mediated by primordial dusty rings and punctuated by episodic accretion is promising and warrants further attention.

At this point we mention some of the limitations of our model. 
The simulations are conducted with the evolution of the magnetic field in the flux-freezing approximation. 
The nonideal MHD effects (Ohmic dissipation, Hall effect, and ambipolar diffusion), which are currently neglected due to high computational cost, may be important in the optically thick and cold environments of PPDs \citep{Lesur+14,Turner+14}.
{  Inclusion of these effects may diminish the magnitude of the magnetic field, thus, reducing $\Sigma_{\rm MRI}$ and increasing the radial extent of the dead zone.
On the other hand, fully resolved MRI turbulence may create electromotive forces that amplify the $z$-component of the magnetic field via an MHD dynamo \citep{Brandenburg95,Gressel10}.
This would work in the opposite direction and reduce the extent of the dead zone.
The diffusion of magnetic field in the innermost regions may also affect the frequency or the shape of the MRI outbursts.
The X-ray luminosity of the accreting star may increase by several orders of magnitude during an FU Orionis outburst \citep{Gudel08, Kastner06}. This may also affect the structure of the dead zone and outburst characteristics.
While the dead zone in our simulations causes a severe bottleneck to the angular momentum transport,
the magnetocentrifugal winds in PPDs may be able to transport angular momentum in the vertical direction and thus affect the structure and evolution of the inner disc \citep{Bai-Stone13a,Whelan+21}.  }
The exact criterion used for planetesimal formation may be a subject of debate, as newer studies suggest marginally different relations \citep[e.g.][]{Li-Youdin21}.
The dust particles cannot grow to large values to enter the Stokes regime in the current model, due to assumptions made in calculating the dust-gas coupling.   
The grown dust also remains in the disc and does not evolve further into larger bodies such as planetary cores or protoplanets, which in turn may perturb the disc \citep{Zhang+18}.

\section{Conclusions}
\label{sec:conclusions}

In this study, we presented the results of numerical experiments conducted using coupled dust-gas global MHD simulations of the long-term evolution of PPDs in the flux-freezing limit. 
The focus of the investigation was on the structure and evolution of primordial dusty rings formed in the inner disc, including the effects of episodic outbursts.
Here we state the salient general findings of this study.
\begin{enumerate}
    \item The dead zone formed with dust+MHD simulations is much more robust in terms of $\alpha_{\rm eff}$ as compared to gas-only simulations. In the dust+MHD simulations, the rings formed within the inner disc are more numerous and span a larger radial extent. This is primarily because solving the detailed ionization-recombination balance equation results in thin MRI active surface layers, leading to diminished $\alpha_{\rm eff}$ and a severe bottleneck to the angular momentum transport.
    \item The conditions in the dusty rings, especially in the innermost regions, are suitable for SI. The dust mass available for planetesimal formation via SI is sufficient to form planetary systems similar to our solar system.
    \item The episodic accretion outbursts significantly perturb the inner disc.
    The dusty rings are accreted on to the central star during an accretion outburst event, which temporarily destroys the SI-prone regions. 
    The planetesimals may start forming during the embedded Class I stage, when the duration between the outbursts starts becoming longer than the timescale of growth for SI.
    The mass available for planetesimal formation increases as the system evolves into the Class II stage.
    \item With an increase in the initial mtf ratio, the planetesimal formation may become less efficient.
    The overall evolution of the disc also depends on the initial cloud core mass.
    An increased core mass results in a greater available mass for planetesimals as well as an increased outburst frequency.
    The disc conditions are especially favourable to SI in a low mass system due to the dust dynamics as well as the scarcity of episodic outbursts, which may explain a higher frequency of terrestrial planetary systems around M-dwarfs.
\end{enumerate}

\section*{Acknowledgements}

{  We thank Indrani Das, Vardan Elbakyan, and Chao-Chin Yang for useful discussions.
We also thank the anonymous referee, whose suggestions improved the quality of the paper.} 
E. I. V. acknowledges support of Ministry of Science and
Higher Education of the Russian Federation under the grant 075-15-2020-780
(N13.1902.21.0039; Sect. 2 and 3). S. B. is supported by a Discovery Grant from NSERC of Canada. 
{ We acknowledge support from the Munich Institute for Astro-, Particle and BioPhysics, funded by the Deutsche Forschungsgemeinschaft (EXC-2094 ? 390783311).} The simulations were performed on the VSC Vienna Scientific Cluster.

\vspace{-0.5 cm}

\section*{Data Availability}

The data underlying this article are obtained with the FEOSAD code and can be shared on reasonable request.

\vspace{-0.5 cm}


\bibliographystyle{mnras}
\bibliography{references} 





\bsp	
\label{lastpage}
\end{document}